\documentclass{sig-alternate}

\usepackage{cite}
\usepackage{graphicx}
\usepackage{amsmath}
\usepackage{subfig}
\usepackage{float}
\usepackage{bm}
\usepackage{xspace}
\usepackage[english]{babel}
\usepackage{ctable}
\usepackage{tabularx}
\usepackage{colortbl}
\usepackage{blindtext}
\usepackage{hyperref}
\usepackage{lastpage}
\usepackage{balance}
\usepackage{mathtools}
\usepackage{multirow}
\usepackage{array}
\usepackage{hhline}

\newcounter{dhecounter}
\DeclareRobustCommand{\dhe}[1]{\textbf{/* #1 (dhe) */}\stepcounter{dhecounter}\typeout{LaTeX Warning: dhe comment \thedhecounter: #1 (line \the\inputlineno)}}
\newcounter{dlacounter}
\DeclareRobustCommand{\dla}[1]{\textbf{/* #1 (dla) */}\stepcounter{dlacounter}\typeout{LaTeX Warning: dla comment \thedlacounter: #1 (line \the\inputlineno)}}
\newcounter{fgecounter}
\DeclareRobustCommand{\fge}[1]{\textbf{/* #1 (fge) */}\stepcounter{fgecounter}\typeout{LaTeX Warning: fge comment \thefgecounter: #1 (line \the\inputlineno)}}
\newcounter{klecounter}
\DeclareRobustCommand{\kle}[1]{\textbf{/* #1 (kle) */}\stepcounter{klecounter}\typeout{LaTeX Warning: kle comment \theklecounter: #1 (line \the\inputlineno)}}
\newcounter{mstcounter}
\DeclareRobustCommand{\mst}[1]{\textbf{/* #1 (mst) */}\stepcounter{mstcounter}\typeout{LaTeX Warning: mst comment \themstcounter: #1 (line \the\inputlineno)}}
\newcounter{swacounter}
\DeclareRobustCommand{\swa}[1]{\textbf{/* #1 (swa) */}\stepcounter{swacounter}\typeout{LaTeX Warning: swa comment \theswacounter: #1 (line \the\inputlineno)}}

\def\term#1{%
	\ifcsname TERM#1\endcsname%
	#1%
	\else%
	\textit{#1}%
	\expandafter\gdef\csname TERM#1\endcsname{}%
	\fi}

\newcommand{\figrow}{-}
\newcommand{\figwidth}{1.}

\newcommand{\FigBiasStrength}{0005}

\newcommand{\fnurl}[2]{%
  \href{#2}{#1}\footnote{\url{#2}}%
}

\newcommand{\bfsection}[1]{\noindent\textbf{#1}}

\newcounter{findingscounter}
\newcommand{\findingbox}[1]{%
	\stepcounter{findingscounter}%
	\vspace{-\baselineskip}
	\begin{center} %
		\framebox{ %
			\begin{minipage}[h]{.95\columnwidth} %
				\textbf{Finding~\thefindingscounter:} #1 %
			\par\xdef\tpd{\the\prevdepth} %
			\end{minipage} %
		}%
	\end{center} %
}

\renewcommand{\findingbox}[1]{ %
	\noindent\textbf{Finding.} #1 %
	}

\usepackage{amssymb}
\usepackage{verbatim}

\begin{document}

\setcopyright{acmcopyright}

\doi{-}

\isbn{-}

\conferenceinfo{-}{-}

\acmPrice{\$15.00}

\conferenceinfo{-}{-}

\title{Assessing the Navigational Effects of Click Biases\\and Link Insertion on the Web
	}

\iftrue
    \numberofauthors{5} %
    \author{
    \alignauthor
    Florian Geigl\\
           \affaddr{KTI, Graz University of Technology}\\
           \email{florian.geigl@tugraz.at}
    \alignauthor
    Kristina Lerman\\
           \affaddr{ISI, University of Southern California}\\
           \email{lerman@isi.edu}
    \and       
    \alignauthor
    Simon Walk\\
           \affaddr{IICM, Graz University of Technology}\\
           \email{simon.walk@tugraz.at}
    \and       
    \alignauthor
    Markus Strohmaier\\
           \affaddr{University of Koblenz-Landau and GESIS}\\
           \email{markus.strohmaier@gesis.org}
    \alignauthor
    Denis Helic\\
           \affaddr{KTI, Graz University of Technology}\\
           \email{dhelic@tugraz.at}
    }
\fi

\date{21 December 2015}

\maketitle

\begin{abstract}
Websites have an inherent interest in steering user navigation in order to, for example, increase sales of specific products or categories, or to guide users towards specific information. In general, website administrators can use the following two strategies to influence their visitors' navigation behavior. First, they can introduce \textit{click biases} to reinforce specific links on their website by changing their visual appearance, for example, by locating them on the top of the page. Second, they can utilize \textit{link insertion} to generate new paths for users to navigate over. 
In this paper, we present a novel approach for measuring the potential effects of these two strategies on user navigation. Our results suggest that, depending on the pages for which we want to increase user visits, optimal link modification strategies vary. Moreover, simple topological measures can be used as proxies for assessing the impact of the intended changes on the navigation of users, even before these changes are implemented.
\end{abstract}

\section{Introduction}
\label{sec:introduction}
Millions of people use the Web on a daily basis to buy products in online shops, perform financial transactions via online banking, or simply browse information systems, media libraries or online encyclopedias, such as IMDb, Netflix or Wikipedia.
To find and access relevant information on the Web, people either search, navigate, or combine these two activities. 
A recent study~\cite{Gleich2010} found that $35\%$ of all visits to a website can be attributed to teleports, which are the direct result of clicks on search-engine results, navigation through manually typed URLs, or clicks on browser bookmarks. The remaining $65\%$ of the clicks can be attributed to the task of navigating a webpage\xspace. In this paper, we direct our attention towards these $65\%$ of actions and tackle the question what potential effects we can expect if we influence the link selection process of website\xspace visitors by simple link modifications. %
In particular, we are interested in the effects of different link modification strategies on (stochastic) models of Web navigation.

\bfsection{Problem.}
By inserting new links between webpages\xspace of a website\xspace, we alter the link structure. This has the potential to change user browsing behavior, since new links create new paths for users to explore the website\xspace. Alternatively, without changing the link structure of the website\xspace, we might be able to influence the link selection process of visitors. Studies have shown that the decisions of users for where to navigate next can be influenced by the layout and the position of the links on a webpage\xspace. In particular, due to position bias~\cite{lerman_pos_bias} users are more likely to select links higher up on webpages\xspace~\cite{murphy2006primacy, buscher2009you, dimi_www_poster}. As a result, inducing click biases, such as repositioning links on a webpage\xspace, highlighting the links, or even making them visually more appealing, can affect the users' decision of where to click next on a website\xspace, similar to the way that adding new links affects browsing.

In this paper we are particularly interested in investigating and comparing the potential consequences of inserting new links and modifying already existing links on the navigational behavior of users. These newly obtained insights are of a significant practical relevance for website owners, as they can be used, for example, by owners of media libraries to increase visits of specific media files in order to reduce the number of different files that need to be cached on fast storage devices. Another example includes online encyclopedias, where operators may want to guide users towards articles of a specific category over some period of time (e.g., the birthday of an inventor). In some of these cases, link insertion might be more time-consuming than simply changing the layout of the website\xspace to increase visibility of specific links and vice versa. Theoretically, we would like to analyze and compare the effects of such link modification endeavors. Practically, new tools are needed to assist website\xspace operators in deciding which of the two strategies they should deploy to achieve the desired effects.

\bfsection{Methods.}
In this paper we study the impact of link modifications on the random surfer, which we apply as a proxy for real user behavior. In the past, a user's decision to click on a link on a webpage\xspace was successfully modeled using the random surfer~\cite{brin, Helic2013, west2012automatic}. In this model, a user selects one of the links on a webpage\xspace uniformly at random and navigates to the page to which the link points. Apart from the huge success of the Google search engine, whose ranking algorithm is based on the random surfer model, empirical studies have shown that this model provides a very precise approximation of real browsing behavior in many situations and for a variety of applications~\cite{brin, geigl_iknow}. An important property of a random surfer is its \textit{stationary distribution}, which is the probability distribution of finding a random surfer at a specific webpage\xspace in the limit of large number of steps.

In particular, we investigate how the random surfer's stationary distribution of a subset of pages (i.e., \textit{target pages}) of a given website\xspace changes as a consequence of (i) modifying already existing links towards them, (ii) introducing new links towards them, or by (iii) combining these two approaches. To that end, we introduce a 
\textit{click bias}, and a \textit{link insertion} strategy. %
We model the effects of click biases on the intrinsic attractiveness of a link to the user by increasing the weight of that link. 
In practice, we may introduce such click biases, for example, by locating the corresponding link on the top of a page. With link insertion, we simply introduce new links between webpages\xspace of a website\xspace, for example, by linking towards a given target page from the starting page.

We introduce quantitative measures that allow us to address the following research questions:

\noindent\textit{Navigational Boost}. How stable is the stationary distribution with respect to the proposed modification strategies, and what are the limits of stationary distributions that can be achieved for a given set of webpages\xspace? Is it (theoretically) possible to achieve a given stationary probability distribution for an arbitrary subset of webpages\xspace of a website\xspace? %
What is the connection between simple topological measures of the website\xspace network and stationary probability?

\noindent\textit{Influence Potential}. 
What is the relative gain of the stationary probabilities compared to their unmodified counterparts. This provides us with an answer to the  ``guidance'' potential of a set of webpages\xspace, defining to what extent it is possible to increase the relative stationary probabilities as compared to the initial unmodified values.

\noindent\textit{Combinations}.
Finally, we are interested how combinations of the two proposed link modification strategies perform in terms of increased stationary probabilities of selected subpages. In particular, we investigate the performance of certain combinations across several different networks and/or selected subpages.

\bfsection{Contributions \& Findings.}
We find that intuitions about how either modification strategy affects navigation are not always correct. Further, our experiments show that the size of a set of targeted subpages is not always a good predictor for the observed effects. Rather, other topological features often better reflect the consequences of a modification. Practically, we provide an \fnurl{open source framework}{ https://github.com/floriangeigl/RandomSurfers} for website\xspace administrators to estimate the effects of link modifications on their website\xspace. %

\section{Related Work}
\label{sec:related_work}
The random surfer model has received much attention from the research community~\cite{lovasz1993random, woess1994random}. While the model is  very simple, it became well-established over the last years. It was applied to a variety of problems from graph generators over graph analysis to modeling user navigation. Furthermore, the model has been applied to calculate structural node properties in large networks. 
HITS~\cite{KleinbergHITS} and PageRank~\cite{brin, PRordertheweb} rank network nodes according to their values in the stationary distribution of the random surfer model. Especially for the later there exists a detailed analysis ranging from the efficiency of its calculation towards its robustness~\cite{insidepr, deepinsidepr}. Bianchini et al.~\cite{insidepr} provided an in-depth analysis of how to tweak the cumulative PageRank of a community of websites. They found that splitting up the content of pages onto more highly interlink pages increases the community's cumulative PageRank---since the community is larger it consists of more pages which are able to trap the random surfer for a longer period of time. Moreover, they suggest to avoid dangling webpages\xspace (i.e., pages without links to other pages). In this paper we are also interested in the sum of the random surfers visit probabilities in a community, however we do not use (i) teleportation as in the PageRank model, and (ii) do not modify the network in its size (i.e., number of pages). On the contrary we modify the transition probabilities of certain links and insert new links into the network. Moreover, since all our datasets are strongly connected, we do not face the problem of unwanted high visit probabilities of usually unimportant pages (i.e., dangling nodes)~\cite{insidepr}.

A random surfer can be  steered towards specific nodes in the network by  increasing the probability of traversing links towards those nodes. This can be accomplished by biasing random surfer's link selection strategy so that it is not uniformly random anymore, but biased towards specific nodes. For instance, in the field of information retrieval Richardson et al.~\cite{richardson2001intelligent} successfully applied biased random surfers to increase the quality of search results compared to those achieved using a simple PageRank. At the same time Haveliwala~\cite{haveliwala2003topic, Haveliwala2002} biased PageRank towards topics retrieved from a search query to rank the query results. Utilizing this technique the results where more accurate than those produced using a single, generic PageRank. Moreover, Gyongyi et al.~\cite{Gyongyi2004} successfully used trust as bias to detect and filter out spam pages of search results.
However, later Al-Saffar and Heileman~\cite{boundsoftopicsenspagerank} showed that biased PageRank algorithms generate a considerable overlap in top results with a simple PageRank. Concerning this problem their main suggestion was to use external biases which do not rely onto the underlying link structure of the network. In our paper we randomly decide towards which nodes we bias the random surfer. This allows us to explore the borders of changes in stationary distributions caused by a bias.

Later Helic et al.~\cite{Helic2013} compared click trails characteristics of stochastically biased random surfers with those of humans. Their conclusion was, that biased random surfers can serve as valid models of human navigation. Further, Geigl et al.~\cite{geigl_iknow} validated this by showing that the result vector of PageRank and clickdata biased PageRank have a strong correlation in an online encyclopedia. This is especially interesting, since it creates the connection of our simulation to real human navigation on the web. Additionally, Lerman and Hogg~\cite{lerman_pos_bias} already showed that it is possible to bias the link selection of users. In particular, they came to the conclusion that users are subject to a \textit{position bias}, making the selection of links higher up on webpages\xspace up to a factor of $3.5$ more likely~\cite{murphy2006primacy, buscher2009you, lerman_pos_bias}. Hence, it is of practical relevance to investigate also the effects of \textit{biases} in the link selection process onto the stationary distribution.

Concerning link insertion there already exists work in literature which makes use of statistical methods to suggest new links in network structures to, for instance, increase the performance of chip architectures~\cite{ogras2006s}. In particular, the authors use a standard mesh and insert long-range links, converting the network into a small-world network. This reduced packet latency results in a major improvement in throughput. Another field of research where link insertion is of interest are recommender systems for social friendship networks~\cite{Li2009, silva2010, Moricz2010, bian2011online}. For example, Xie et al.~\cite{Xie2010} characterized interests of users in two dimensions (i.e., context and content) and exploited this information to efficiently recommend potential new friends in an online social network. 
In this paper we focus on the effects of inserted links onto the typical whereabouts of the random surfer. In particular, we are interested in inserting links into the network such that the random surfer more frequently visits a predefined subset of pages of a website\xspace (i.e., target pages).

\section{Methodology}
\label{sec:methodology}
\begin{figure*}[t!]
	\centering
	\captionsetup[subfigure]{justification=centering}
	\newcommand{\markelem}[1]{\textcolor{red}{\bm{#1}}}
	\newcommand{\markmod}[1]{\textcolor{blue}{\bm{#1}}}
	\newcommand{\markstep}[1]{\textbf{#1}}
	\newcommand{\rowtitle}[1]{\rotatebox{90}{\textbf{#1}}}

	\newcolumntype{C}[1]{>{\centering\arraybackslash}m{#1px}}
	\newcommand{\netcolwidth}{45}
	\newcommand{\netwidth}{1.}
	\newcommand{\netspaceing}{\vspace{4px}}
	\begin{tabular}{C{5} C{45} llll}
		\centering
		\vspace{4px}\rowtitle{Original} &
		\multicolumn{1}{C{45}}{\vspace{4px}\includegraphics[width=1.\linewidth]{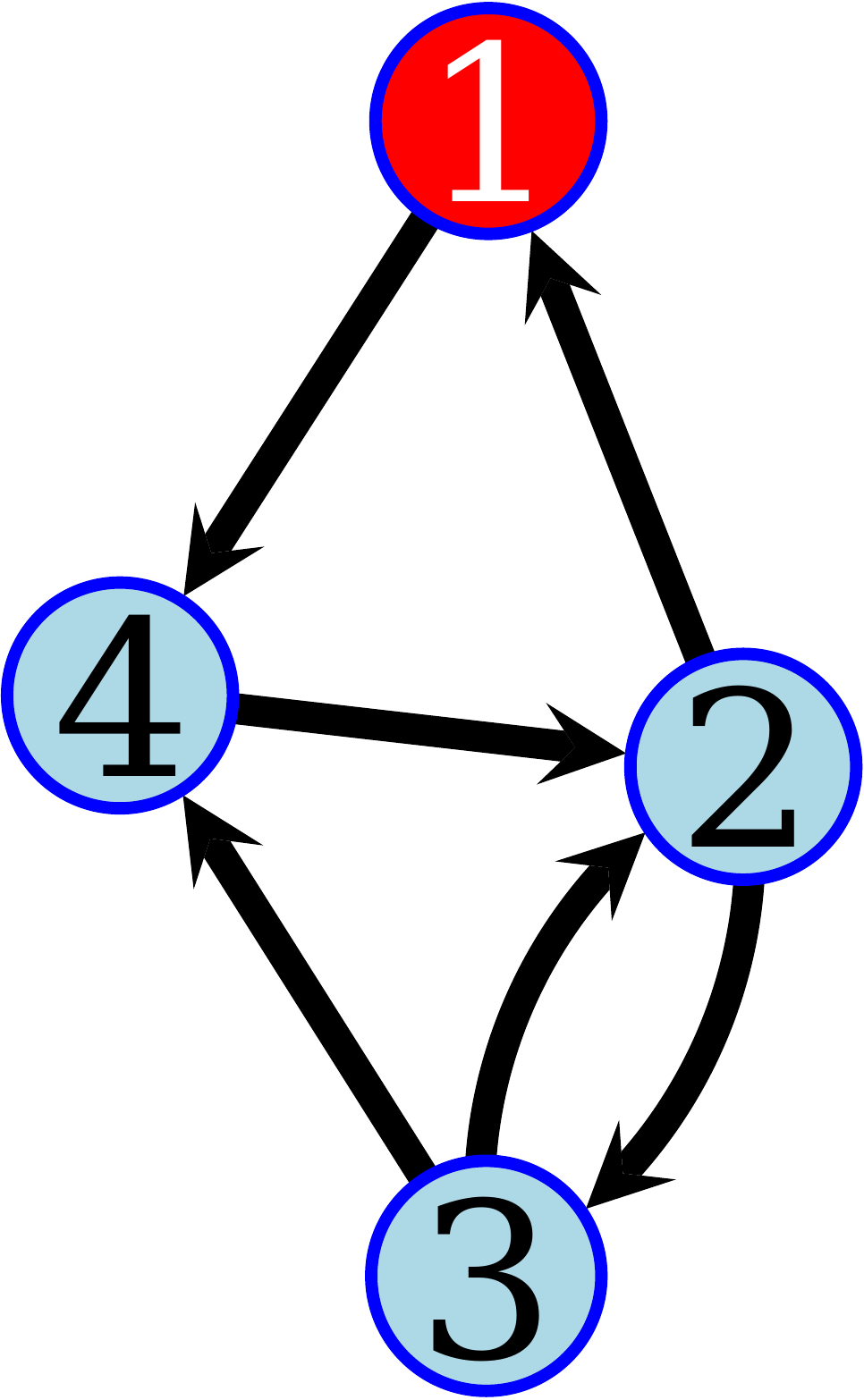}} & &
		$\bm{W} = %
			\begin{pmatrix} 
			0 & 1 & 0 & 0 \\
			0 & 0 & 1 & 1 \\
			0 & 1 & 0 & 0 \\
			1 & 0 & 1 & 0
			\end{pmatrix} $ &
		$\bm{P} = %
			\begin{pmatrix}
			0.0 & 0.5 & 0.0 & 0.0 \\
			0.0 & 0.0 & 0.5 & 1.0 \\
			0.0 & 0.5 & 0.0 & 0.0 \\
			1.0 & 0.0 & 0.5 & 0.0
			\end{pmatrix}
		$ & 
		$\bm{\pi} = %
			\begin{pmatrix}
			\markelem{0.18} \\
			0.36 \\
			0.18 \\
			0.27
			\end{pmatrix}$
		\\ \hline
		\vspace{4px}\rowtitle{Click Bias} &	 \multicolumn{1}{C{45}}{\vspace{4px}\includegraphics[width=1.\linewidth]{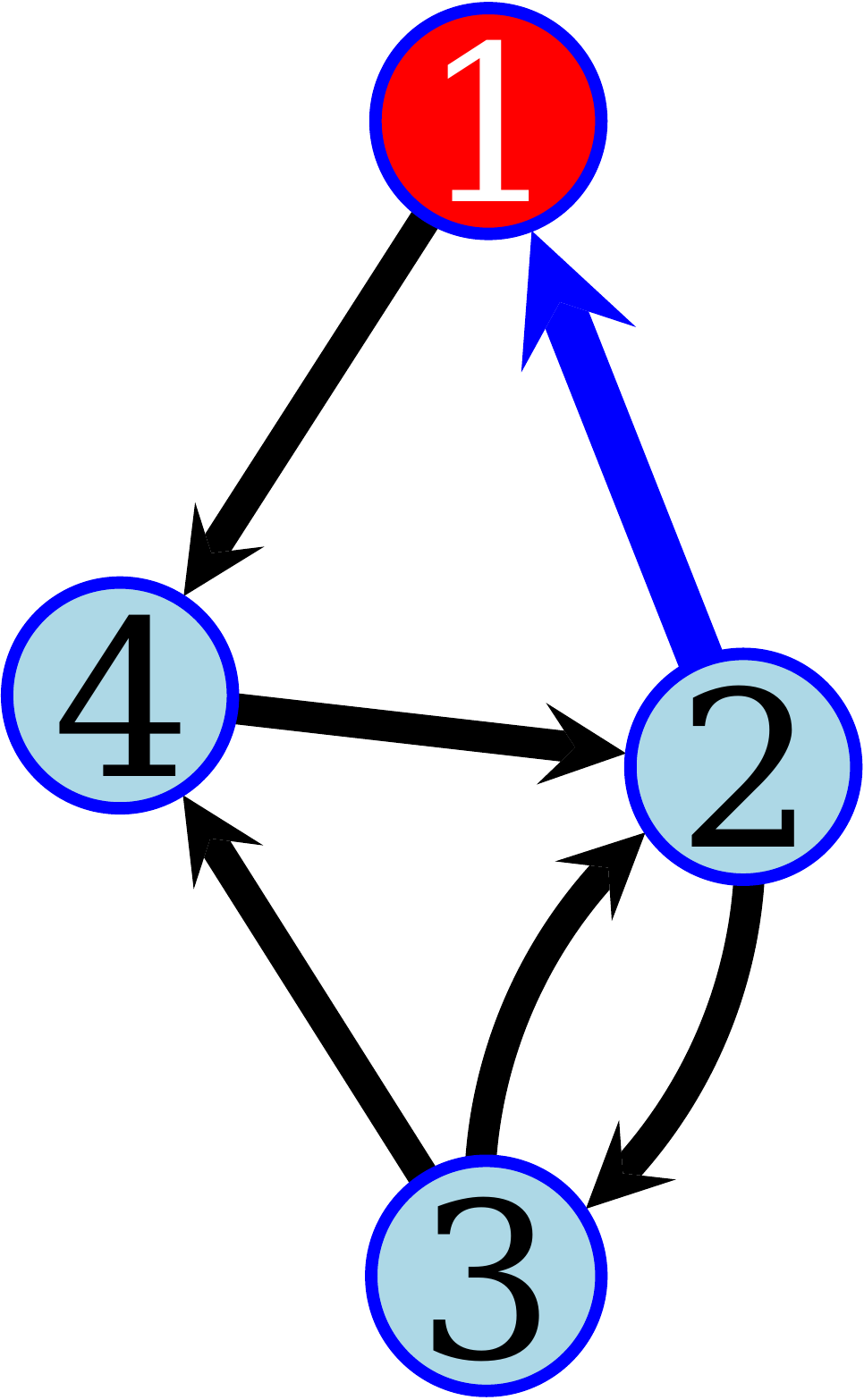}} & 
		$ \bm{B} = %
			\begin{pmatrix}
			\markmod{2} & 0 & 0 & 0 \\
			0 & 1 & 0 & 0 \\
			0 & 0 & 1 & 0 \\
			0 & 0 & 0 & 1
			\end{pmatrix} $ &
		$ \bm{W'} = %
			\begin{pmatrix}
			0 & \markmod{2} & 0 & 0 \\
			0 & 0 & 1 & 1 \\
			0 & 1 & 0 & 0 \\
			1 & 0 & 1 & 0
			\end{pmatrix} $ &
		$ \bm{P'} =
			\begin{pmatrix}
			0.0 & \markmod{0.67} & 0.0 & 0.0 \\
			0.0 & 0.0 & 0.5 & 1.0 \\
			0.0 & \markmod{0.33} & 0.0 & 0.0 \\
			1.0 & 0.0 & 0.5 & 0.0
			\end{pmatrix} $  & 
		$\bm{\pi'} = %
			\begin{pmatrix}
			\markelem{0.24} \\
			0.35 \\
			0.12 \\
			0.29
			\end{pmatrix}$
		\\ \hline
		\vspace{4px}\rowtitle{Link Insertion} &
		\multicolumn{1}{C{45}}{\vspace{4px}\includegraphics[width=1.\linewidth]{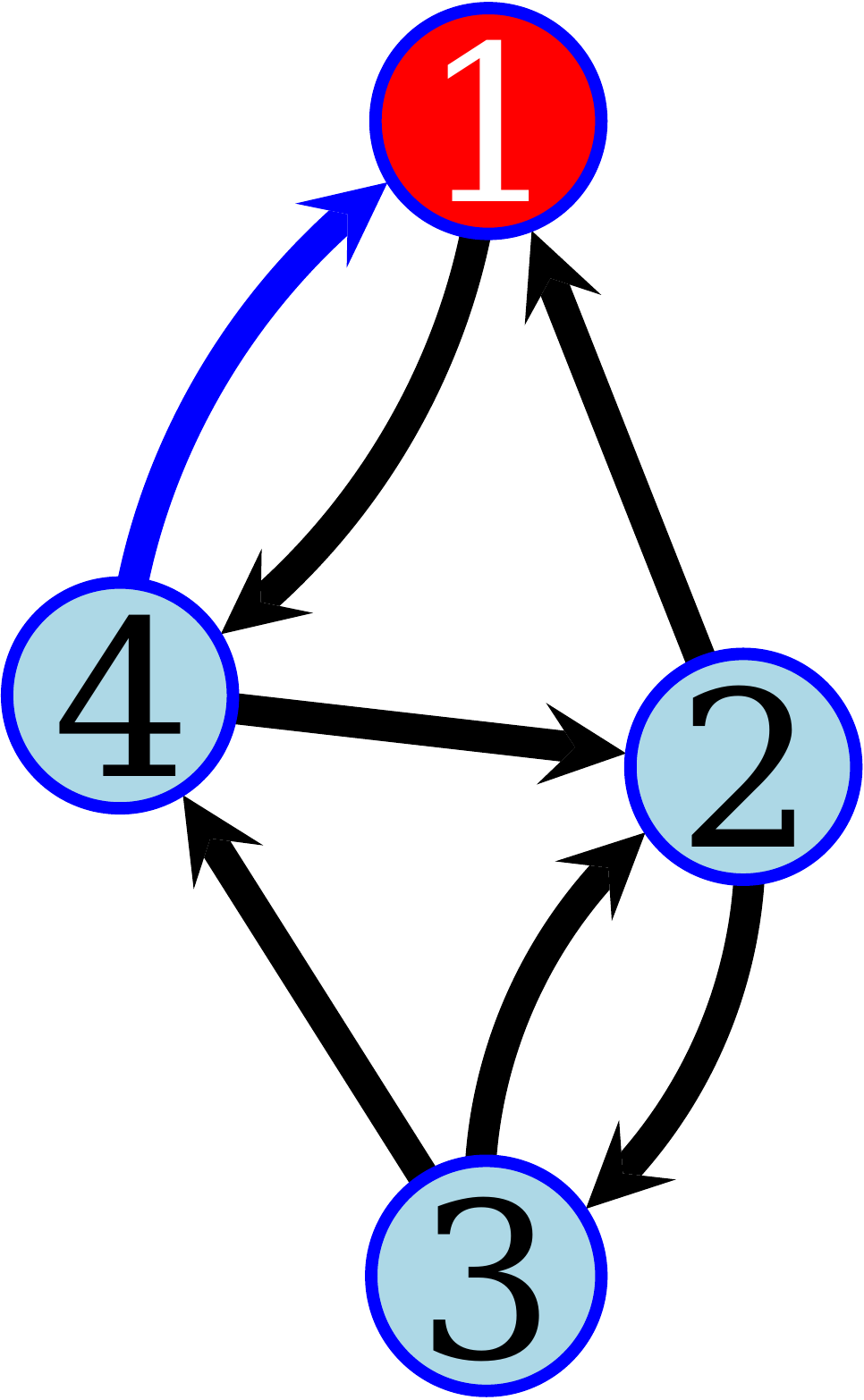}} & &
		$ \bm{W'} = %
			\begin{pmatrix}
			0 & 1 & 0 & \markmod{1} \\
			0 & 0 & 1 & 1 \\
			0 & 1 & 0 & 0 \\
			1 & 0 & 1 & 0
			\end{pmatrix} $ &
		$ \bm{P'} = %
			\begin{pmatrix}
			0.0 & 0.5 & 0.0 & \markmod{0.5} \\
			0.0 & 0.0 & 0.5 & \markmod{0.5} \\
			0.0 & 0.5 & 0.0 & 0.0 \\
			1.0 & 0.0 & 0.5 & 0.0
			\end{pmatrix} $ &
		$\bm{\pi'} = %
			\begin{pmatrix}
			\markelem{0.22} \\
			0.33 \\
			0.17 \\
			0.28
			\end{pmatrix}$
	\end{tabular}
	\DeclareRobustCommand{\toytarnodesvec}{\begin{pmatrix} %
			\markelem{1} & 0 & 0 & 0 %
		\end{pmatrix}^\intercal}
	\caption{\textbf{Modeling Click Bias and Link Insertion - Illustrative Example.} We intend to use different link modification strategies to  steer the random surfer towards the red colored node $1$ more often. Hence, our target nodes vector consists of only one node: node $1$ ($\bm{t} = \toytarnodesvec$). In each row we visualize the corresponding network of the website\xspace, where nodes represent webpages\xspace and links represent hyperlinks between the webpages\xspace. Further, for each of them we show how we calculate its stationary distribution. This involves (from left to right) the weighted adjacency matrix $\bm{W}$ (unmodified network) or $\bm{W'}$ (modified networks), the corresponding transition matrix $\bm{P}$ (unmodified network) or $\bm{P'}$ (modified networks) and finally the corresponding stationary distribution $\bm{\pi}$ (unmodified network) or $\bm{\pi'}$ (modified networks). The blue links in the graphs and the blue matrix elements in bold show the link modifications and their effects on the adjacency and the transition matrix. The red vector elements show the effects of the modifications on the stationary probability (energy) of node $1$. \textbf{Top row.}  Here we depict the original and unmodified network. \textbf{Middle row.} We modify the network with a \textit{click bias}. We double the statistical weights of links towards target nodes (bias strength $b=2$). To calculate the modified adjacency matrix we first construct the diagonal bias matrix $\bm{B}$ and then compute $\bm{W'} = \bm{B} \bm{W}$. We see an increase in energy of node $1$ from $0.18$ in the unmodified network to $0.24$. \textbf{Bottom row.} We insert a new link from node $4$ to $1$ (i.e., blue link in graph and blue element in $\bm{W'}$) into the original network. Due to the link insertion the energy of node $1$ increases from $0.18$ in the unmodified network to $0.22$ in the modified network. Thus, in this toy example the effects of the click bias are stronger than those of link insertion. Additionally, we see that also elements in the out-component of node $1$ (i.e., node $4$) profit of an increased energy of node $1$ since a significant amount of $1$'s increased energy flows into node $4$.}
	\label{fig:edu}
	\vspace*{-1.2\baselineskip}
\end{figure*}

We base our methodology on the calculations of the stationary distribution of a random surfer on the original and manipulated networks. The networks consist of nodes, which represent webpages\xspace and directed links between nodes, which represent hyperlinks between webpages\xspace. We first calculate the transition matrix and the stationary distribution for the original network, which will be used as a baseline for comparing the effects of link modifications. Second, we increase the statistical weight of a random surfer visiting a set of predefined nodes (i.e., \textit{target pages} or \textit{target nodes}). We do that either by increasing the link weights towards selected nodes (click bias) or by adding new links pointing towards those nodes (link insertion). Third, we generate the corresponding transition matrix for the modified network. Fourth, we calculate the stationary distribution of the new transition matrices. Finally, we compare the modified stationary distribution with the original stationary distribution to gain insights into the effects of the different link modifications. Figure~\ref{fig:edu} illustrates these steps on a toy example.

\subsection{Preliminaries}
In what follows we formalize our approach algebraically. We represent a website\xspace as a directed network with a weighted adjacency matrix $\bm{W} \in \mathbb{R}^{n\times n}$, where $n$ is the number of webpages\xspace in the website\xspace under investigation. We define the element $W_{ij}$ of the weighted adjacency matrix $\bm{W}$ as the sum of edge weights of all links pointing from node $j$ to node $i$. For example, $W_{ij}=1$ if there is a single link from page $j$ to page $i$ with weight $1$, and $W_{ij}=3$ if there are three links pointing from page $j$ to page $i$ each with weight $1$.

For our analysis we introduce \textit{target nodes} as the nodes whose stationary probability we want to increase. We use vector $\bm{t} \in \mathbb{R}^{n}$ to specify them:
\begin{equation}
	\label{eq:t}
	t_i = \begin{cases}
	1 &\text{if }i\text{ is a \textit{target node}} \\
	0 &\text{otherwise}.
	\end{cases}
\end{equation}
We further define $\phi$ as a fraction of target nodes with respect to the total number of nodes $n$:
\begin{equation}
	\phi = \frac{\sum_i t_i}{n}
\end{equation}
Hence, $\phi=0.1$ means that $10\%$ of nodes from the network are target nodes. 

\subsection{Stationary Distribution}
The  stationary distribution represents the probability to find the random surfer on any node in the limit of large number of steps. To compute the stationary distribution we first need to construct a diagonal out-degree matrix $\bm{D}$, with the weighted node out-degrees on its diagonal. Using $\text{ diag}(\bm{v})$ to denote diagonal matrices with elements of a vector $\bm{v}$ on their diagonal we define $\bm{D}$ as:
\begin{equation}
	\label{eq:D}
	\bm{D} = \text{ diag}\left(\sum_{i=1}^{n}W_{ij}\right)\,.
\end{equation}
Using $\bm{D}$ matrix we can calculate the transition matrix $\bm{P}$, which is a left stochastic matrix of $\bm{W}$ as $\bm{P}=\bm{W}\bm{D}^{-1}$. The stationary distribution $\bm{\pi}$ now satisfies the (right) eigenvalue equation for the matrix $\bm{P}$: $\bm{\pi}=\bm{P\pi}$. 

\subsection{Click Bias}
To introduce click biases that influence the link selection strategy of the random surfer, we reweigh the links pointing towards target nodes by multiplying their weight by a constant scalar $b$, which we call bias strength. For example, a bias strength of $b=2$ doubles the weight of all links towards target nodes. The final probability of the random surfer to traverse a link is then directly proportional to its weight.

Algebraically, we induce biases with a diagonal bias matrix $\bm{B}$ which we define as $\bm{B} = \bm{I} + (b-1) \cdot \text{ diag}(\bm{t})$. The adjacency matrix of a biased network is $\bm{W'} = \bm{B}\bm{W}$. To compute the stationary distribution of the biased network,  we first calculate the new transition matrix $\bm{P'}=\bm{W'}\bm{D'}^{-1}$ and then its stationary distribution $\bm{\pi'}$.

Please note that from the technical perspective, inducing a bias is the same as inserting parallel links towards target nodes---it increases the value of specific elements (i.e., those representing links towards target nodes) in the adjacency matrix. The total weight of newly added parallel links $l(b)$ due to an induced bias $b$ is given by:
\begin{equation}
	\label{eq:l(s)}
	l(b) = \underbrace{\sum_{ij} W'_{ij}}_{\text{\# links in $\bm{W'}$}} - \underbrace{\sum_{ij} W_{ij}}_{\text{\# links in $\bm{W}$}}
\end{equation}
To allow for a fair comparison between the click bias and the link insertion strategy we insert exactly $l(b)$ new links with weight $1$ in the latter case. %

\subsection{Link Insertion}
\label{sec:meth_link_insertion}
The second link modification strategy consists of inserting new links towards the target nodes from a given set of source nodes. This strategy represents the case where a website\xspace administrator inserts links towards target nodes from important subpages of their website\xspace. We define the importance of a webpage\xspace as its stationary probability in the original network.

To insert a given number $l(b)$ of new links we proceed as follows. We start by sorting nodes by their stationary probability in a descending order. In the next step we insert new links from the top $l(b)/(n\cdot\phi)$ nodes to all target nodes. Here $n\cdot\phi$ is the number of target nodes and we always \textit{ceil} the calculated number of source nodes to ensure that there are enough pairs of nodes. If one of the target nodes is itself designated as a source node we do not insert self-loops---from the practical point of view, it does not make sense to link a webpage\xspace to itself. In the rare case where we have connected all possible combinations of source and target nodes but did not reach the required number of links, we simply reiterate the list of the source nodes resulting in parallel links between nodes. Please note that we insert parallel links if a link between a source and a target node has already existed in the original network. However, this happens extremely rarely because all of our networks are sparse. In fact, in all our experiments the fraction of inserted parallel links was on average less than $1\%$.

\subsection{Combinations}
Finally, we can combine the two link modification strategies and study the effects of such combinations on the stationary distribution and investigate if an optimal combination of strategies exists, which outperforms the individual approaches. From the practical point of view this means that for optimally steering website\xspace users, we combine both, the click bias and link insertion mechanisms.

To create a combined link modification method we first introduce $\alpha \in [0,1]$, which we call the mixing factor. The mixing factor determines how many of the $l(b)$ links are inserted by the click bias. Then, $1-\alpha$ defines how many links are inserted by the link insertion strategy:
\begin{equation}
	\label{eq:convex_combination}
	l(b) = \underbrace{\alpha \cdot l(b)}_{\text{\# biased links}} + \underbrace{(1-\alpha) \cdot l(b)}_{\text{\# inserted links}}
\end{equation}
With a combined strategy we cannot bias all links towards target nodes---again, we need to select a subset of links towards target nodes. In analogy to the link insertion method we again preferably select links between nodes having higher stationary probability in the unmodified network. Thus, we first compute the probability distribution over the eligible links in the form of matrix $\bm{L}$, where $\sum_{ij}L_{ij}=1$. We define matrix $\bm{L}$ as:
\begin{equation}
	\bm{L} = \text{ diag}(\bm{\pi}) \cdot \text{ diag}(\bm{t}) \cdot \bm{W} \cdot \text{ diag}(\bm{\pi}).
\end{equation}
The probability of selecting a link is directly proportional to the product of the unmodified stationary probability of its source and target node. Note that due to the multiplicative factor $\text{ diag}(\bm{t}) \cdot \bm{W}$ only links towards target nodes have a non-zero probability. With $\bm{L}$ in place we sample $\alpha \cdot l(b)$ links without replacement and multiply their value in $\bm{W'}$ by $b$ to induce the click bias. To insert the remaining $(1-\alpha) \cdot l(b)$ links we adopt the link insertion strategy on the matrix $\bm{W'}$ as described previously.

\subsection{Measuring the Effects}
\label{sec:methodology_measurements}
To measure the effects of link modification strategies we quantify how the stationary probabilities of given target nodes change as a function of the modification. In the remainder of this paper we will refer to a node's stationary probability using the, in the literature established, term \textbf{\textit{energy}}~\cite{insidepr}. To that end, we calculate the \textit{energy of target nodes} ($\pi'_t$), which is the sum of the modified stationary probabilities of target nodes, as following:
\begin{equation}
    \label{eq:sum_stat_dist_group}
    \pi'_t = \sum_{i} \pi'_i \cdot t_i \, ,
\end{equation}
where $\bm{\pi'}$ is the stationary distribution of the modified adjacency matrix. 

We further measure the \textit{influence potential}, which is the relative increase in the energy of  target nodes due to the modification, as a factor $\tau$:
\begin{equation}
	\label{eq:stat_prob_fac}
	\tau = \frac{\pi'_t}{\pi_t}, 
\end{equation} 
where $\pi_t$ is the energy of target nodes of the unmodified network (i.e., $\pi_t = \sum_i \pi_i \cdot t_i$).

\section{Datasets}
\label{sec:datasets}
For our experiments we use three datasets: an online encylopedia \fnurl{Wikipedia for Schools}{http://http://schools-wikipedia.org/} (\term{W4S}\xspace) and two online media libraries \fnurl{ORF TVthek}{http://tvthek.orf.at/} (\term{ORF}\xspace) and \fnurl{Das Erste Mediathek}{http://mediathek.daserste.de/} (\term{DEM}\xspace). 

We collected the data by crawling the corresponding websites\xspace. Starting from the main page of a website\xspace we recursively crawled all subpages by following all outgoing links from a given webpage\xspace. Note that we did not follow external links, meaning that we skipped links to pages not belonging to a given website\xspace. Further, we did not follow links generated via Flash, AJAX or any other client-rendered content. 

After collecting the data, we removed self-loops, which are links from a webpage\xspace to itself, and  special links such as ``log-in'', ``write a review'', and all other links that require a session-id. In the next step, we represented each dataset as a  directed network---webpages\xspace are represented as nodes connected by directed links. For calculating the stationary distribution, we extracted the largest strongly connected component (SCC) of each network, so that in the final network it is possible to navigate from any given node to any other node in the network. These final networks have $4,051$ nodes and $111,795$ links (\textit{\term{W4S}\xspace}), $9,799$ nodes and $301,844$ links (\textit{\term{ORF}\xspace}), and $70,063$ nodes and $3,448,513$ links (\textit{\term{DEM}\xspace}).

\section{Experimental Setup}
To investigate the effects of manipulating links we first generate sets of target nodes. For this purpose we draw the desired number of nodes uniformly at random from the network without replacement, creating a synthetic set of nodes of a specified size. Note that those sets can consist of webpages\xspace that are not linked to each other. We conduct all of our experiments with the same initially generated target nodes to reduce the influence of the random node selection process. For making the number of webpages\xspace selected as target nodes comparable between datasets we refer to the size of target nodes as $\phi$, which is the fraction of target nodes. To generate target nodes we use several values for $\phi$ which range from $0.01$ to $0.2$. %
For each dataset and each $\phi$ we generate $100$ different synthetic sets of nodes (i.e., target nodes).

\bfsection{Limiting (High) Bias Behavior.}
In our first experiment we are interested in analyzing the impact of an increasing bias strength on the energy of target nodes using either a click bias on already existing links or inserting new links in an informed way. We use bias strengths reaching from $b=2$ to $b=200$ to investigate their effects. Note that for the link insertion strategy the number of inserted links is defined by the bias strength $b$ using Equation~\ref{eq:l(s)}. This ensures a fair comparison between the two methods.

\bfsection{Realistic (Lower) Bias Strengths.}
In this experiment we investigate practically relevant~\cite{lerman_pos_bias, hoggDisentangling} values for the bias strength $b$. In particular, we iterate over the range $2$ to $15$ as bias strengths. With this experiment we gain insights into the effects of the proposed modifications, which can be implemented in websites\xspace. After the modification of the adjacency matrix we measure the energy of target nodes $\pi'_t$. This allows us to investigate the efficiency of both methods for a given bias strength.

\bfsection{Relative Increase in Stationary Probability.}
With the previous experiments we analyze changes in the energy of target nodes in absolute terms. For instance, we may learn that for a given set of target nodes we may achieve an energy of $\pi'_t=0.5$. However, we do not know what the relative increase in their energy is. For example, the set of target nodes may have had $\pi_t=0.49$ in the unmodified network rendering our efforts futile in \textit{relative} terms. Thus, in this experiment
we use $\tau$ to measure the influence potential. A higher value for $\tau$ means a larger \textit{relative} increase in the energy of target nodes. Again, we compare the results for a given bias strength between our two methods.

\bfsection{Combination of Strategies.}
Finally, we are interested in investigating if and to what extent the energy of target nodes changes if we combine click biases and link insertion. We vary the mixture factor $\alpha$ from $0$ to $1$ in increments of $0.1$ and measure the energy of target nodes $\pi'_t$ of the modified networks.

\section{Results \& Discussion}
\label{sec:results}

\subsection{Saturation}
\label{sec:results:saturation}

\begin{figure}[t!]
	\captionsetup[subfigure]{justification=centering}
	\captionsetup[subfloat]{farskip=2pt,captionskip=1pt}
	\centering
	\renewcommand{\figwidth}{.48\linewidth}
	\includegraphics[width=.8\linewidth]{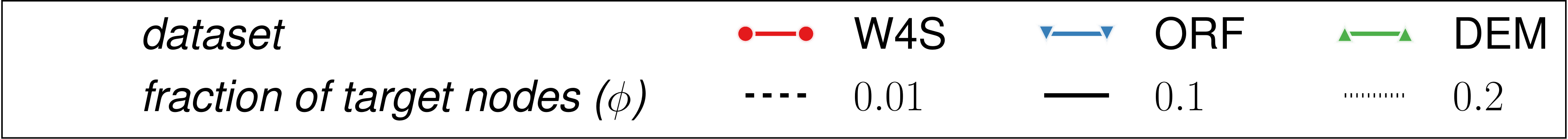}
	\hspace*{\fill}
	\subfloat[Click Bias]{\includegraphics[width=.48\linewidth]{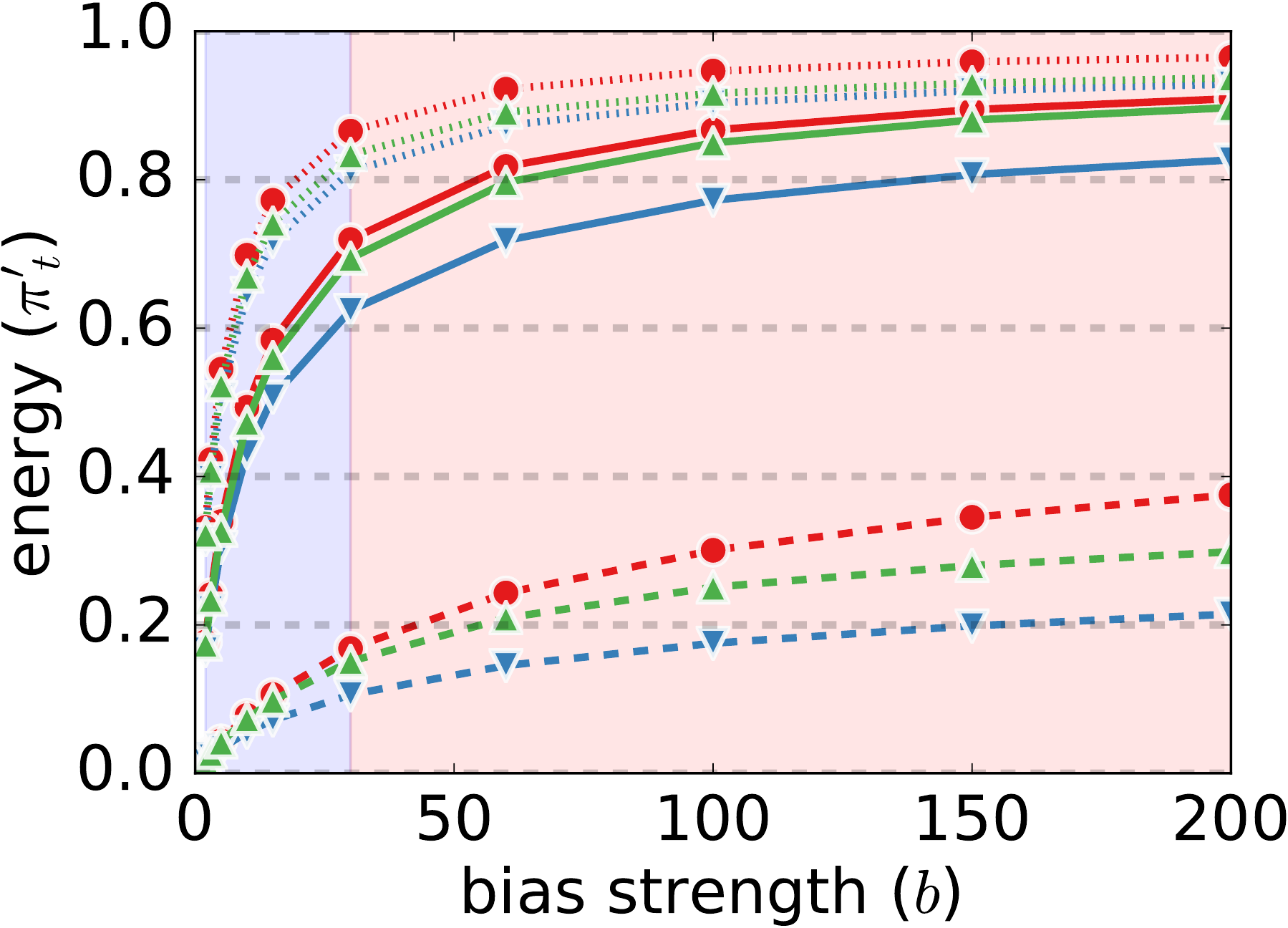}} \hfill
	\subfloat[Link Inseration]{\includegraphics[width=.48\linewidth]{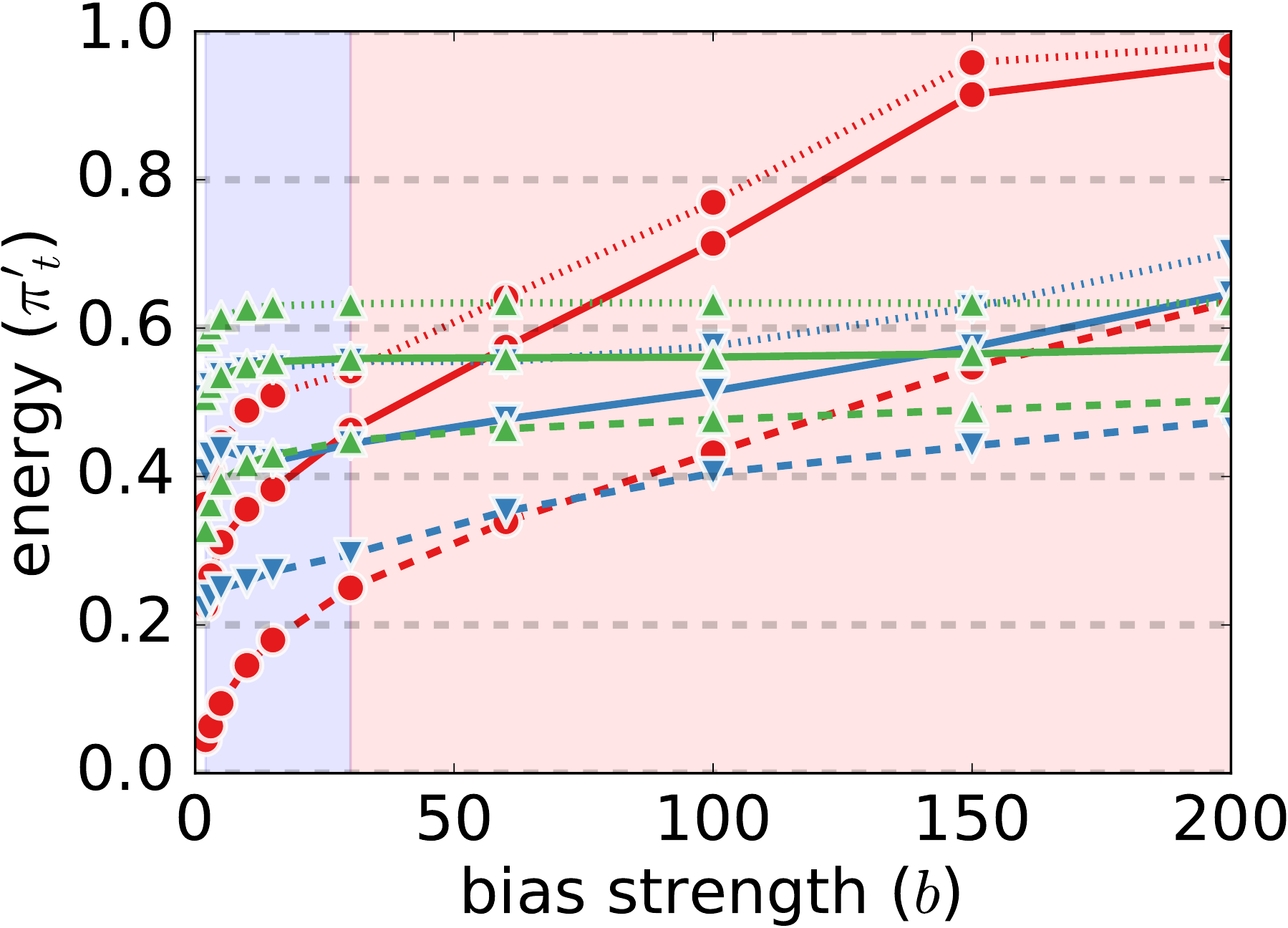}\label{fig:stat_prob:LI}} \hfill
	\vspace*{-.8\baselineskip}
	\caption{\textbf{Saturation.} The plots depict the connection between bias strength (x-axis) and the increased energy of target nodes due to an induced click bias (left) or link insertion (right). Each marker type and color refers to one dataset. Dashed, solid and dotted line styles refer to fraction of target nodes $\phi$ $0.01$, $0.1$ and $0.2$ respectively. We can observe that both link modification strategies reach a certain level of saturation---meaning that further increases in bias strength do not result in an increase in energy of target nodes. Therefore, for both strategies we identify two phases: a (i) \textit{navigational boost} phase in which we observe a rapid increase of the stationary probability (blueish region with small values of the bias strength), and a (ii) \textit{saturation} phase (reddish region with larger values of the bias strength).}
	\label{fig:saturation}
	\vspace*{-1.2\baselineskip}
\end{figure}

\begin{figure*}[ht!]
	\renewcommand{\FigBiasStrength}{0150}
	\newcommand{\method}{_link_ins}
	\captionsetup[subfigure]{justification=centering}
	\captionsetup[subfloat]{farskip=2pt,captionskip=1pt}
	\centering	
	\includegraphics[width=.4\linewidth]{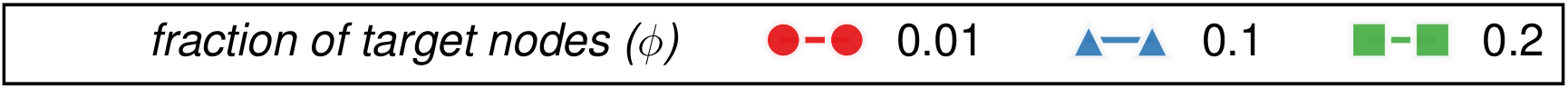}	
	\renewcommand{\figwidth}{.3\linewidth}	
	
	\iffalse
	\renewcommand{\figrow}[2]{
		\subfloat[#1 in-degree]{\includegraphics[width=.3\linewidth]{#2_bs0150_com_in_deg_lines}}\hfill
		\subfloat[#1 out-degree]{\includegraphics[width=.3\linewidth]{#2_bs0150_com_out_deg_lines}}\hfill
		\subfloat[#1 degrees-ratio]{\includegraphics[width=.3\linewidth]{#2_bs0150_ratio_com_out_deg_in_deg_lines}}
	}
\subfloat[\term{W4S}\xspace in-degree]{\includegraphics[width=.3\linewidth]{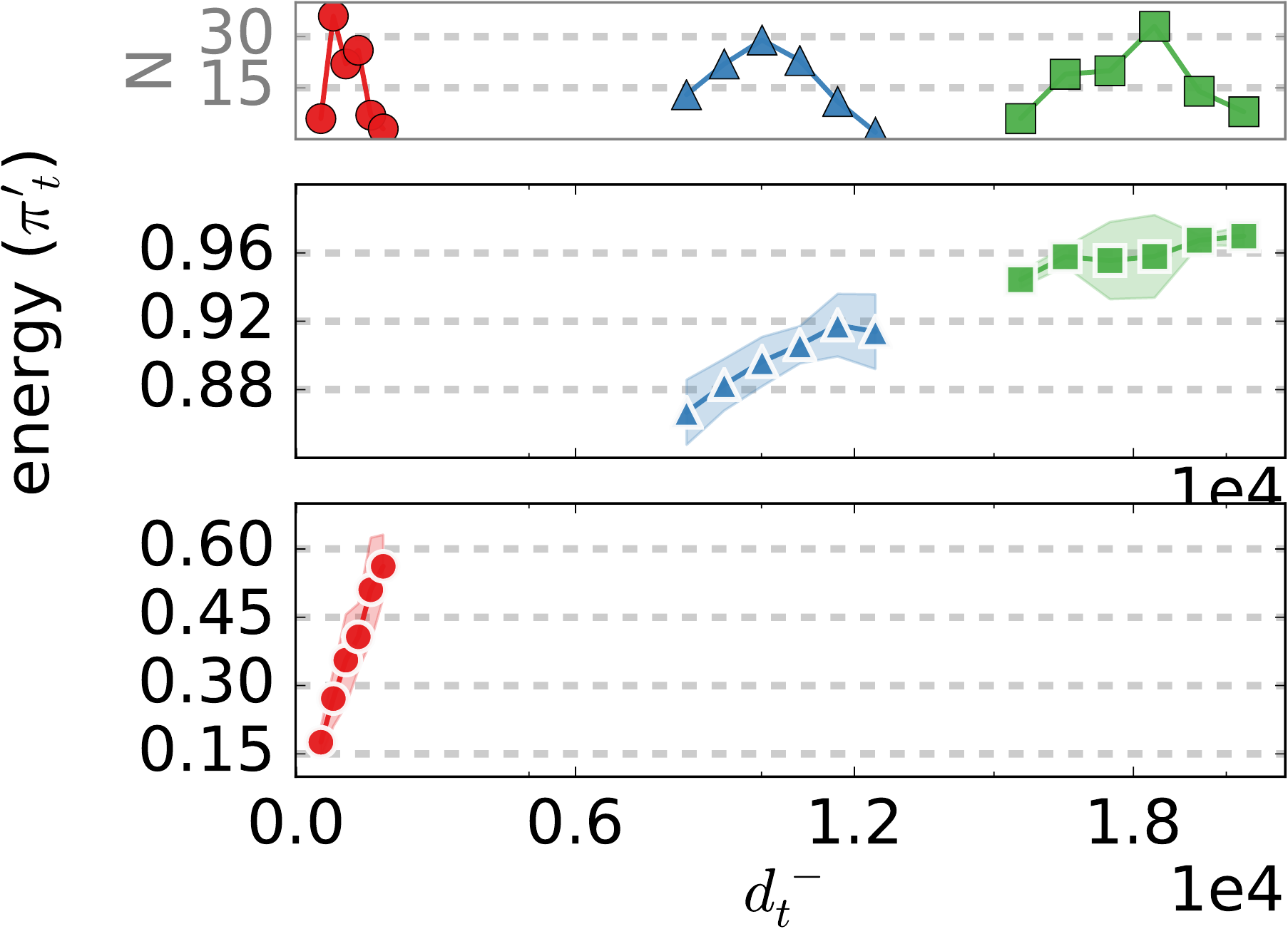}}\hfill
\subfloat[\term{W4S}\xspace out-degree]{\includegraphics[width=.3\linewidth]{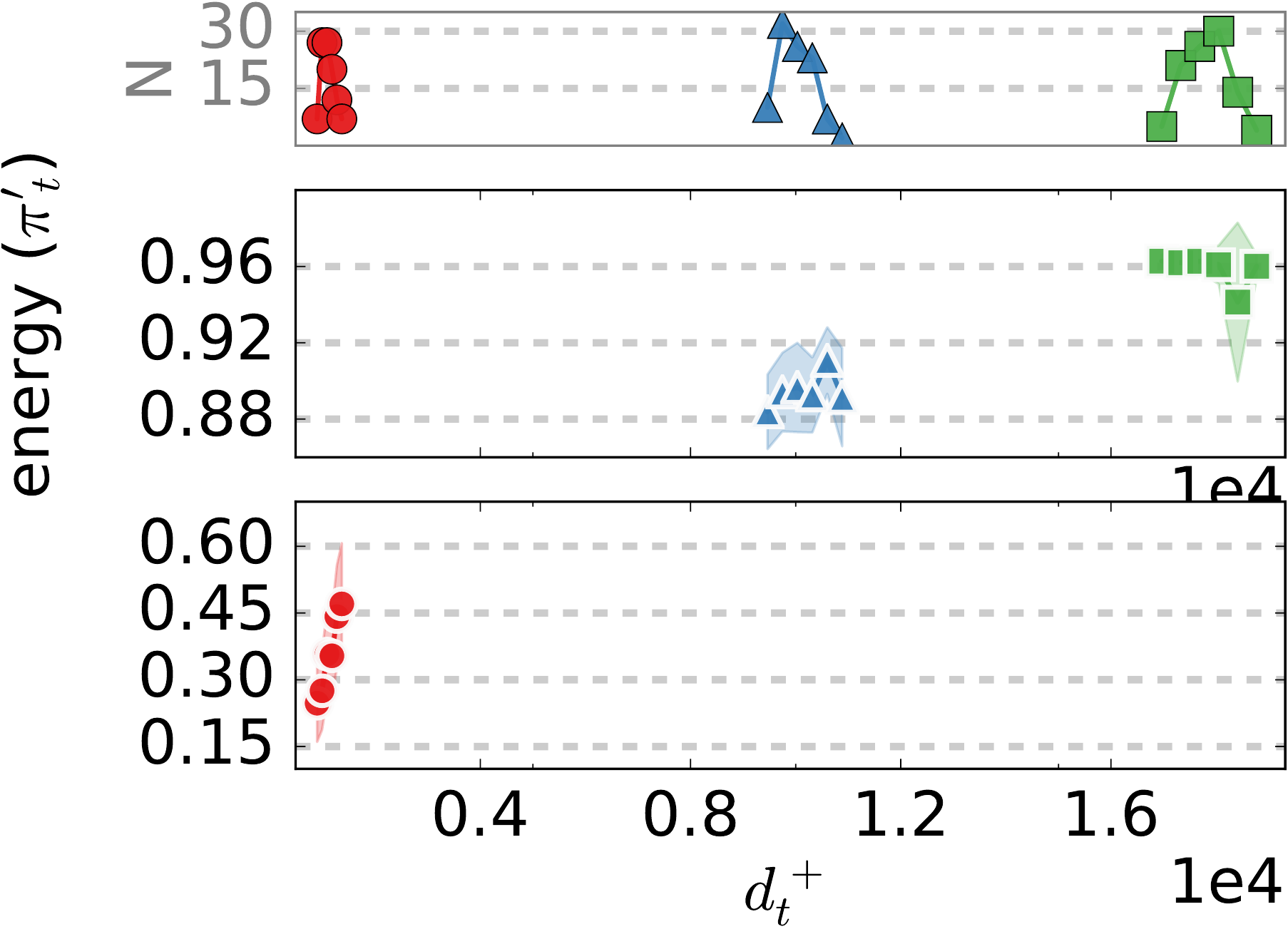}}\hfill
\subfloat[\term{W4S}\xspace degrees-ratio]{\includegraphics[width=.3\linewidth]{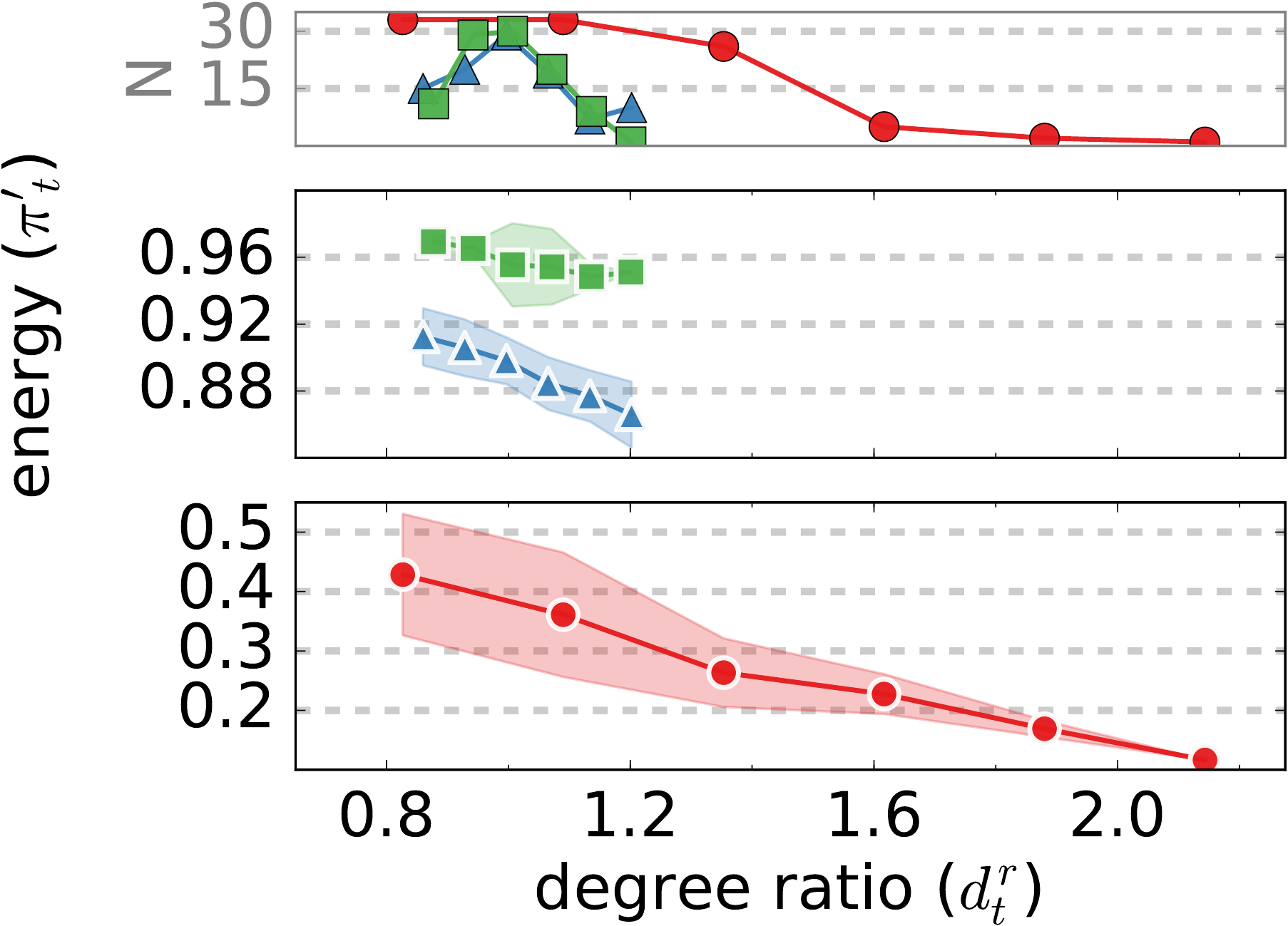}}
	\vspace{-.6\baselineskip}
\subfloat[\term{ORF}\xspace in-degree]{\includegraphics[width=.3\linewidth]{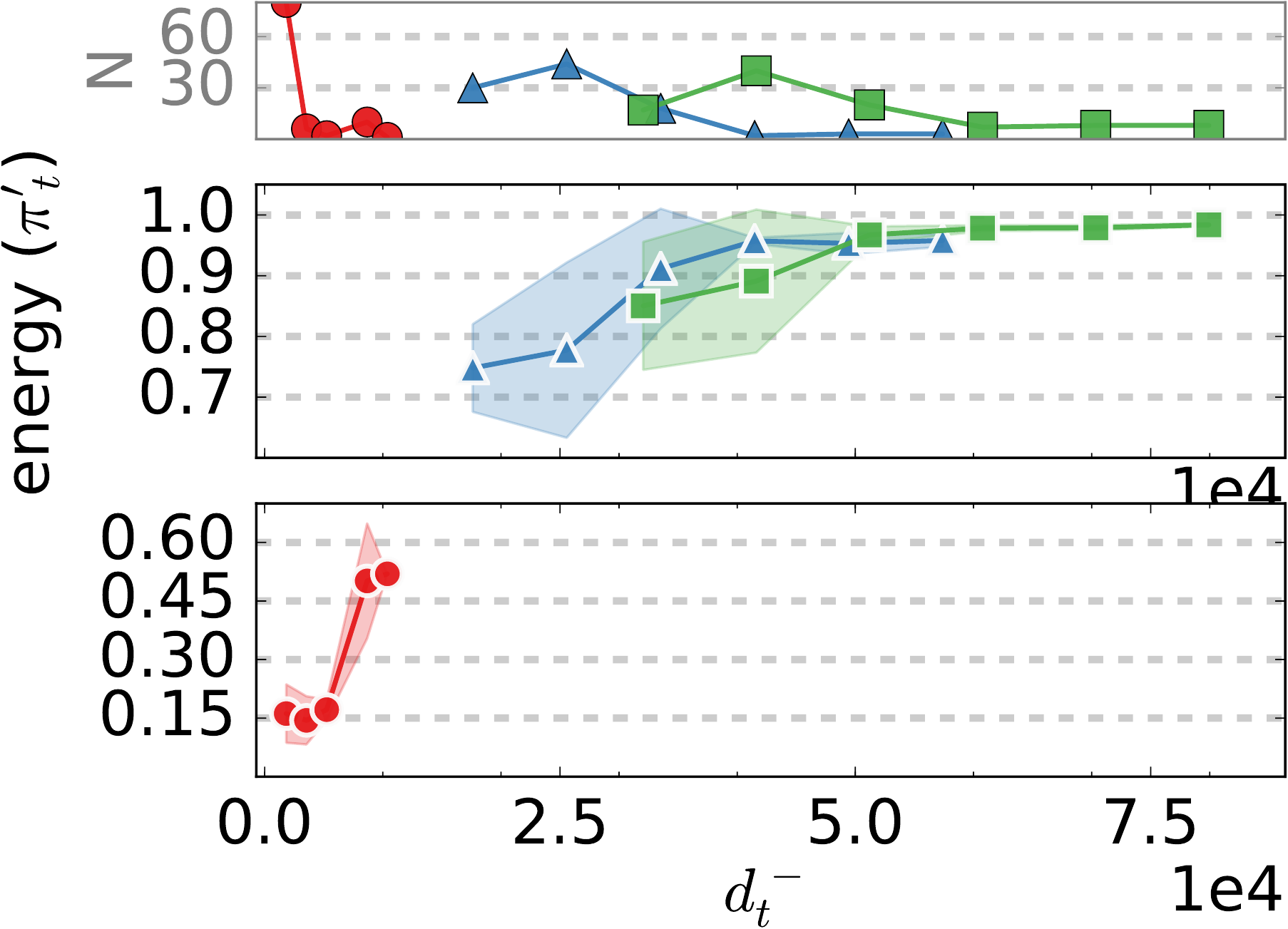}}\hfill
\subfloat[\term{ORF}\xspace out-degree]{\includegraphics[width=.3\linewidth]{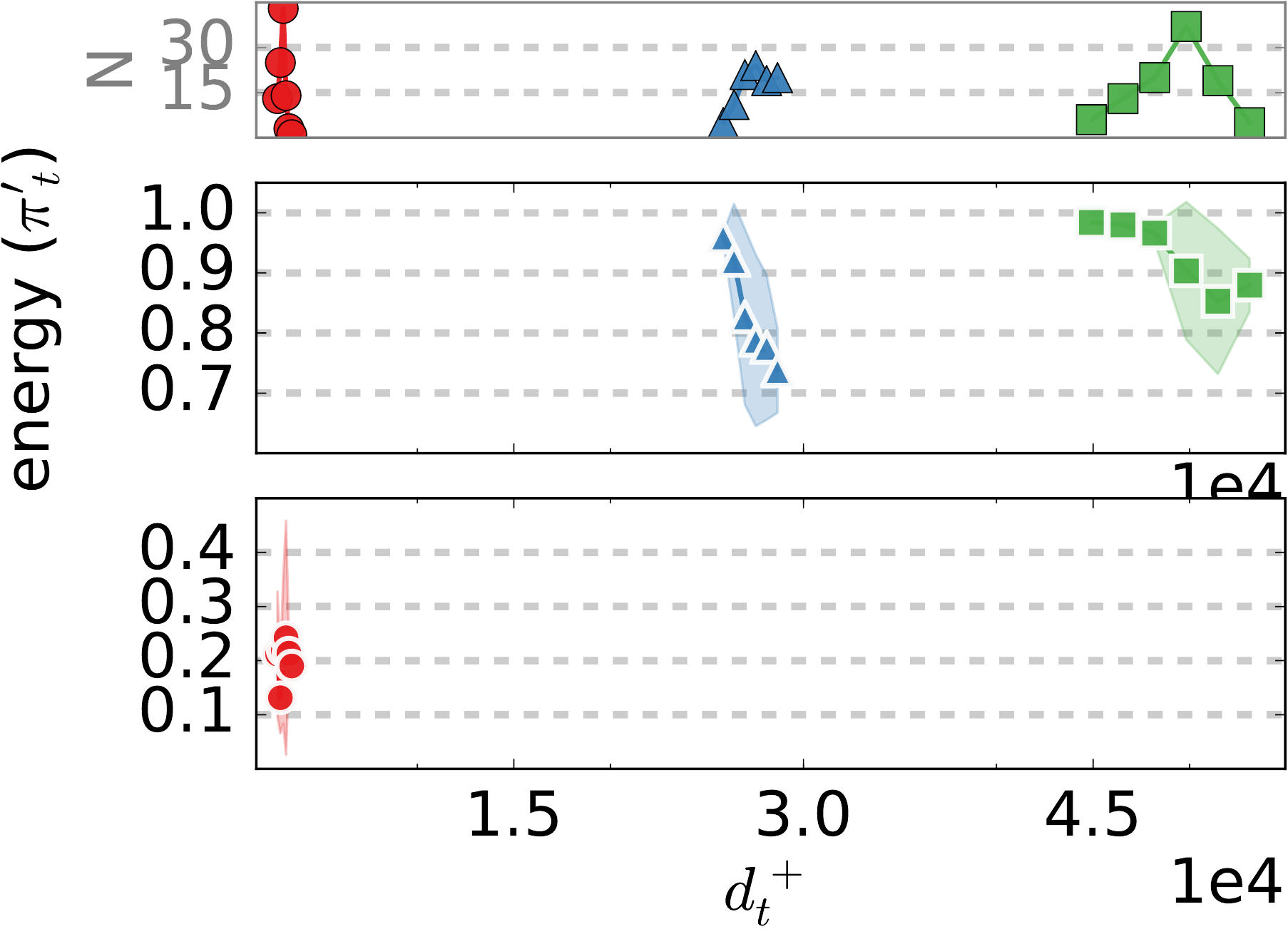}}\hfill
\subfloat[\term{ORF}\xspace degrees-ratio]{\includegraphics[width=.3\linewidth]{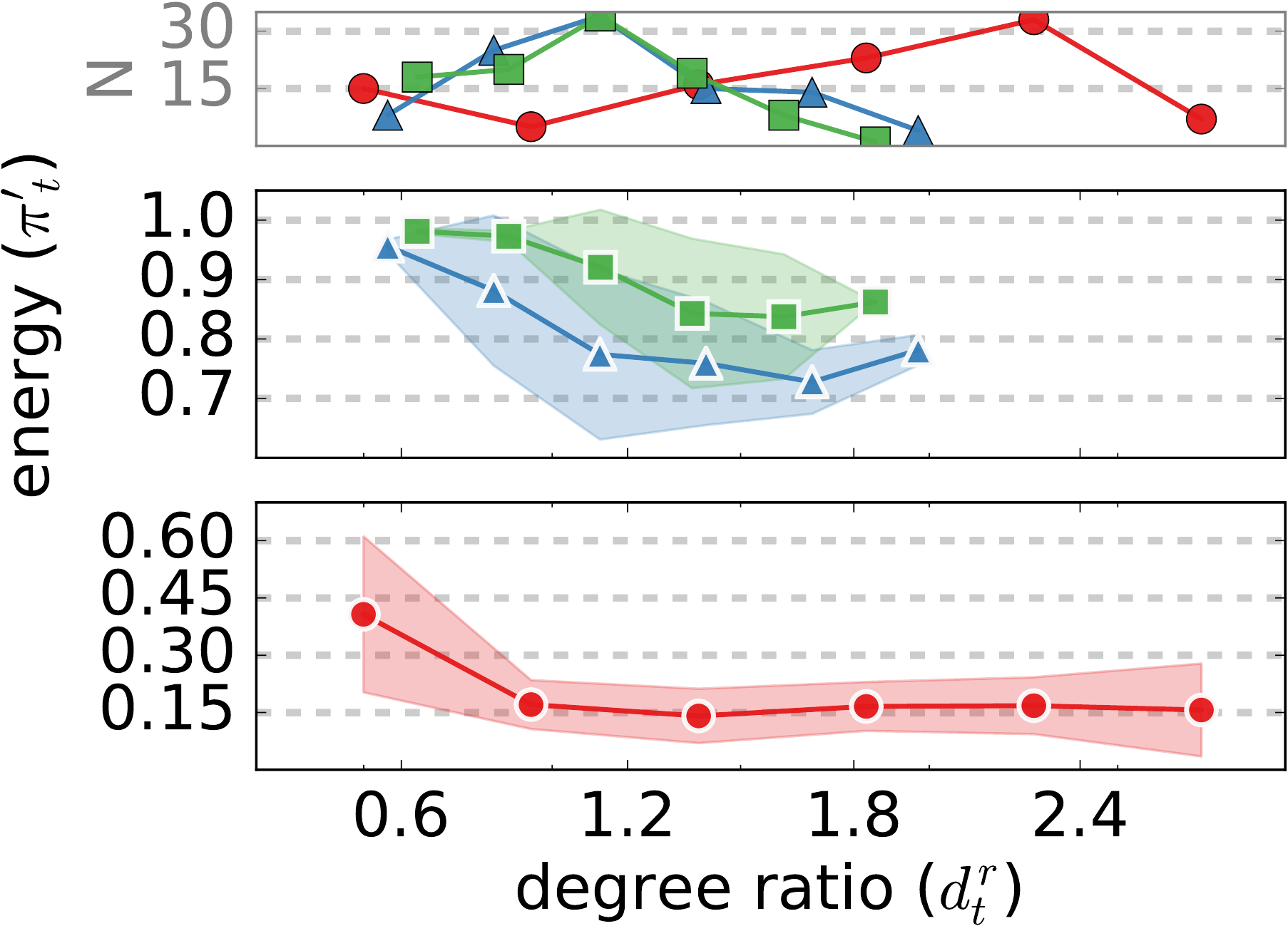}}
	\vspace{-.6\baselineskip}
\subfloat[\term{DEM}\xspace in-degree]{\includegraphics[width=.3\linewidth]{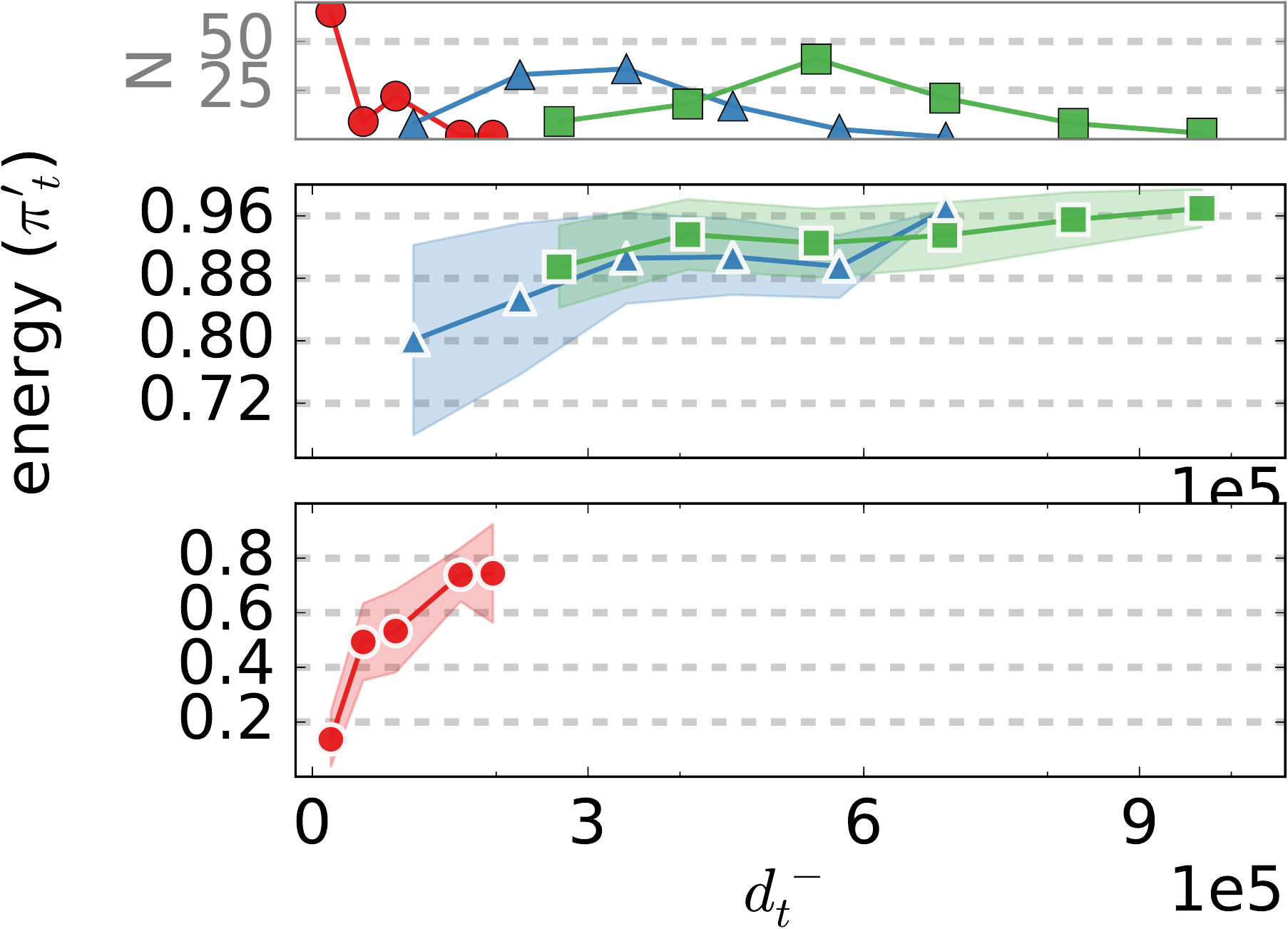}}\hfill
\subfloat[\term{DEM}\xspace out-degree]{\includegraphics[width=.3\linewidth]{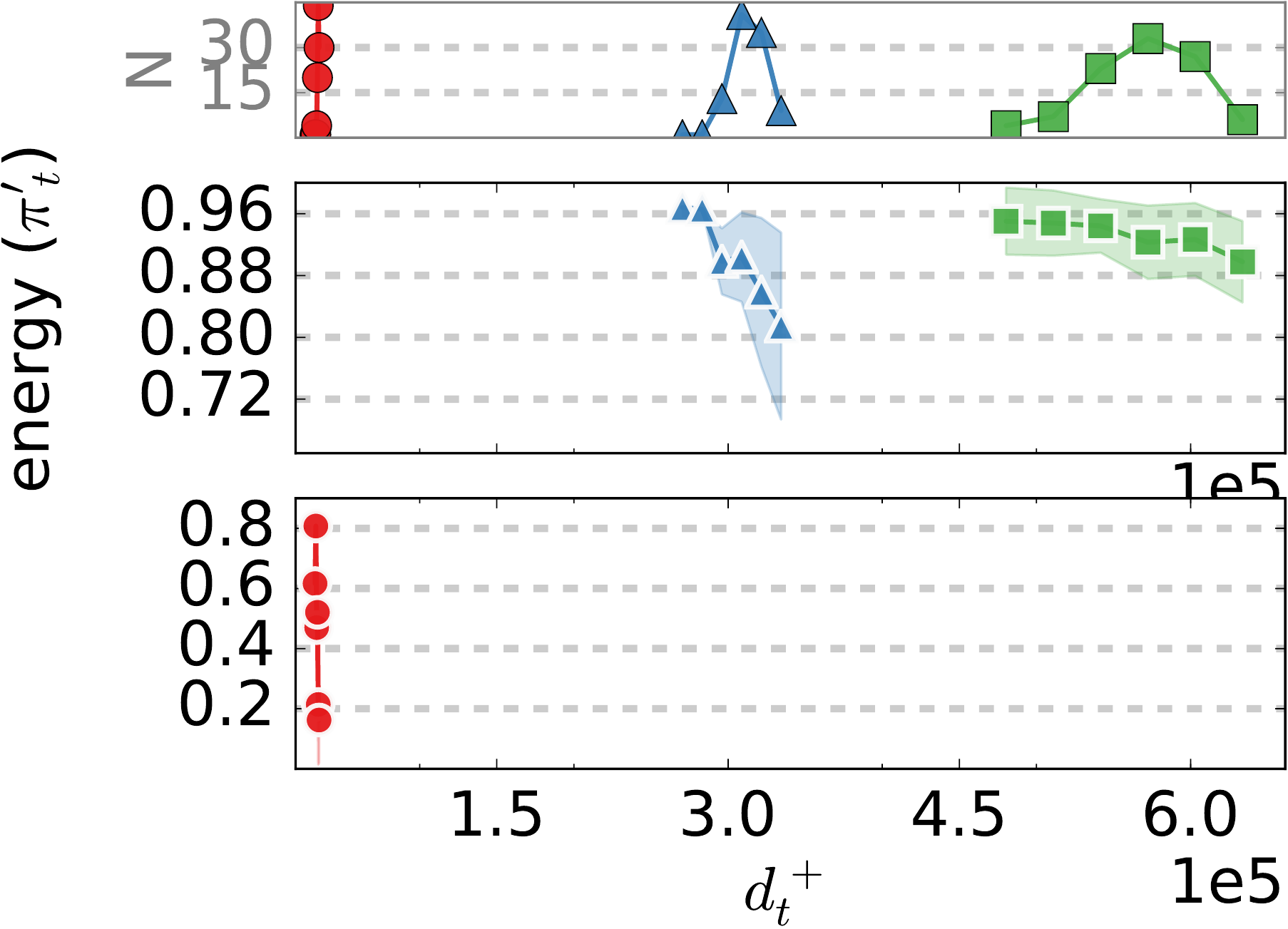}}\hfill
\subfloat[\term{DEM}\xspace degrees-ratio]{\includegraphics[width=.3\linewidth]{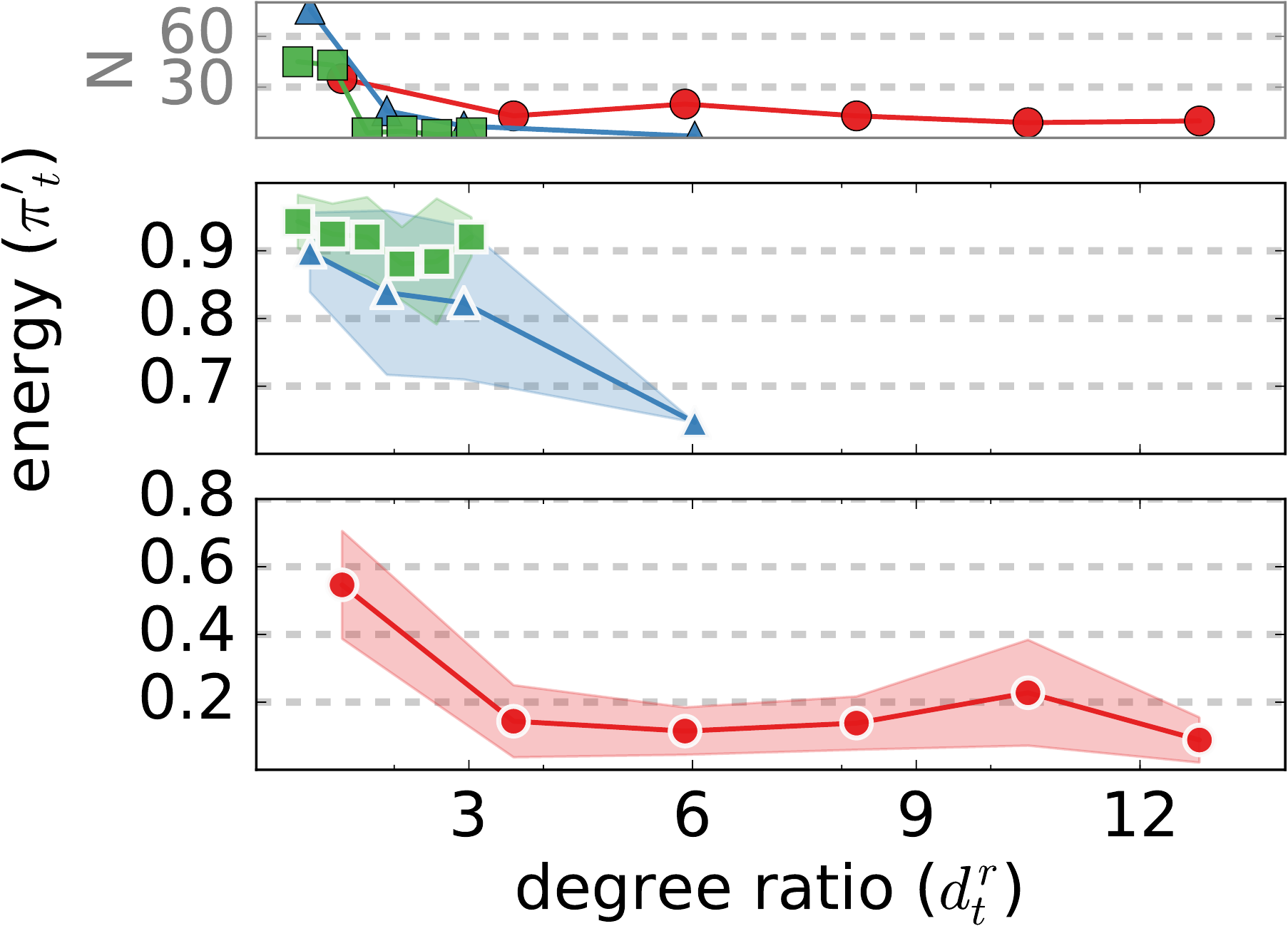}}
	\else
	\renewcommand{\figrow}[3]{
		\hspace*{\fill}
		\subfloat[\term{W4S}\xspace]{\includegraphics[width=.3\linewidth]{#1_bs0150_ratio_com_out_deg_in_deg_lines_link_ins}}\hfill
		\subfloat[\term{ORF}\xspace]{\includegraphics[width=.3\linewidth]{#2_bs0150_ratio_com_out_deg_in_deg_lines_link_ins}}\hfill
		\subfloat[\term{DEM}\xspace]{\includegraphics[width=.3\linewidth]{#3_bs0150_ratio_com_out_deg_in_deg_lines_link_ins}}
		\hspace*{\fill}
	}
\hspace*{\fill}
\subfloat[\term{W4S}\xspace]{\includegraphics[width=.3\linewidth]{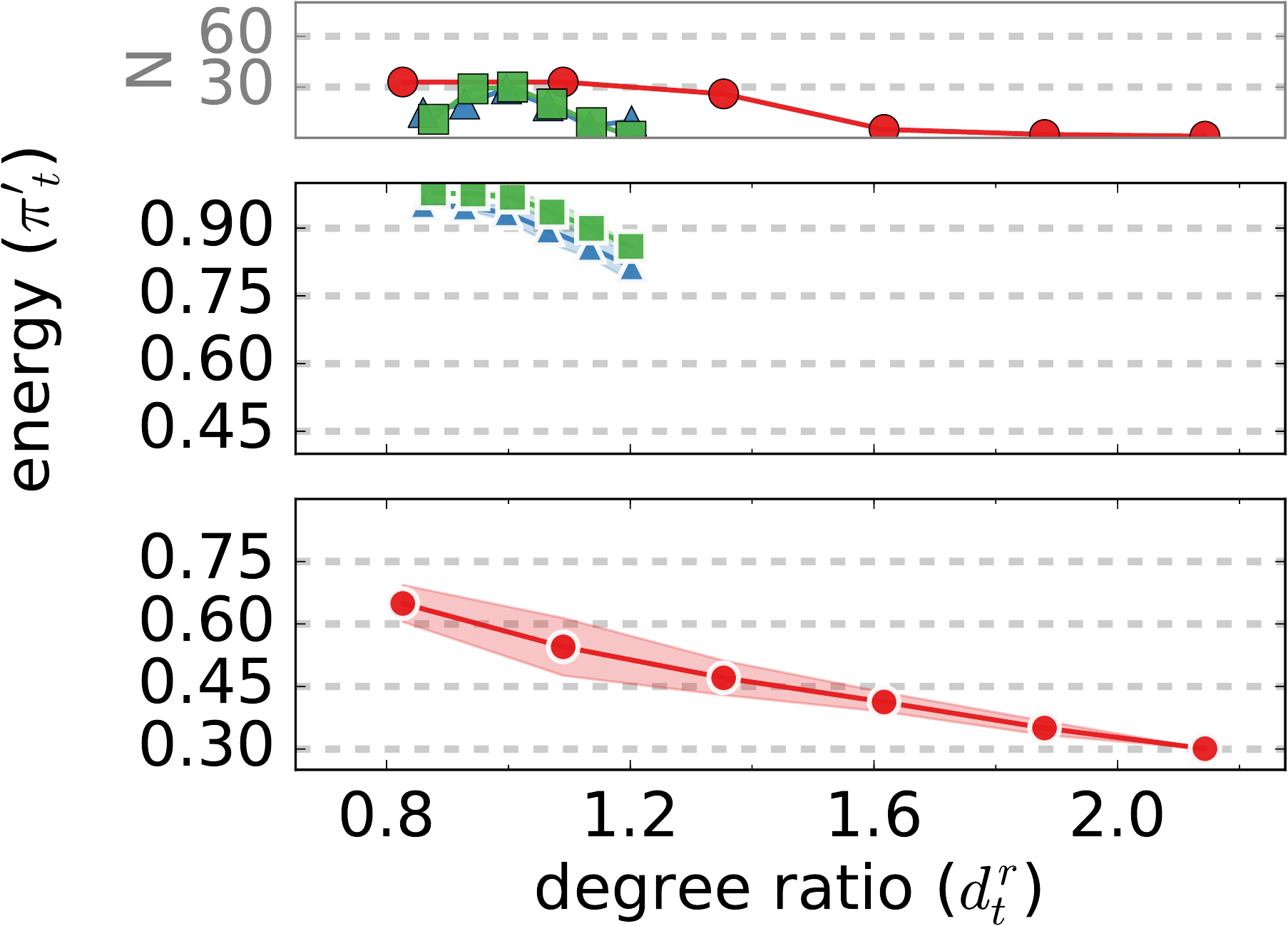}}\hfill
\subfloat[\term{ORF}\xspace]{\includegraphics[width=.3\linewidth]{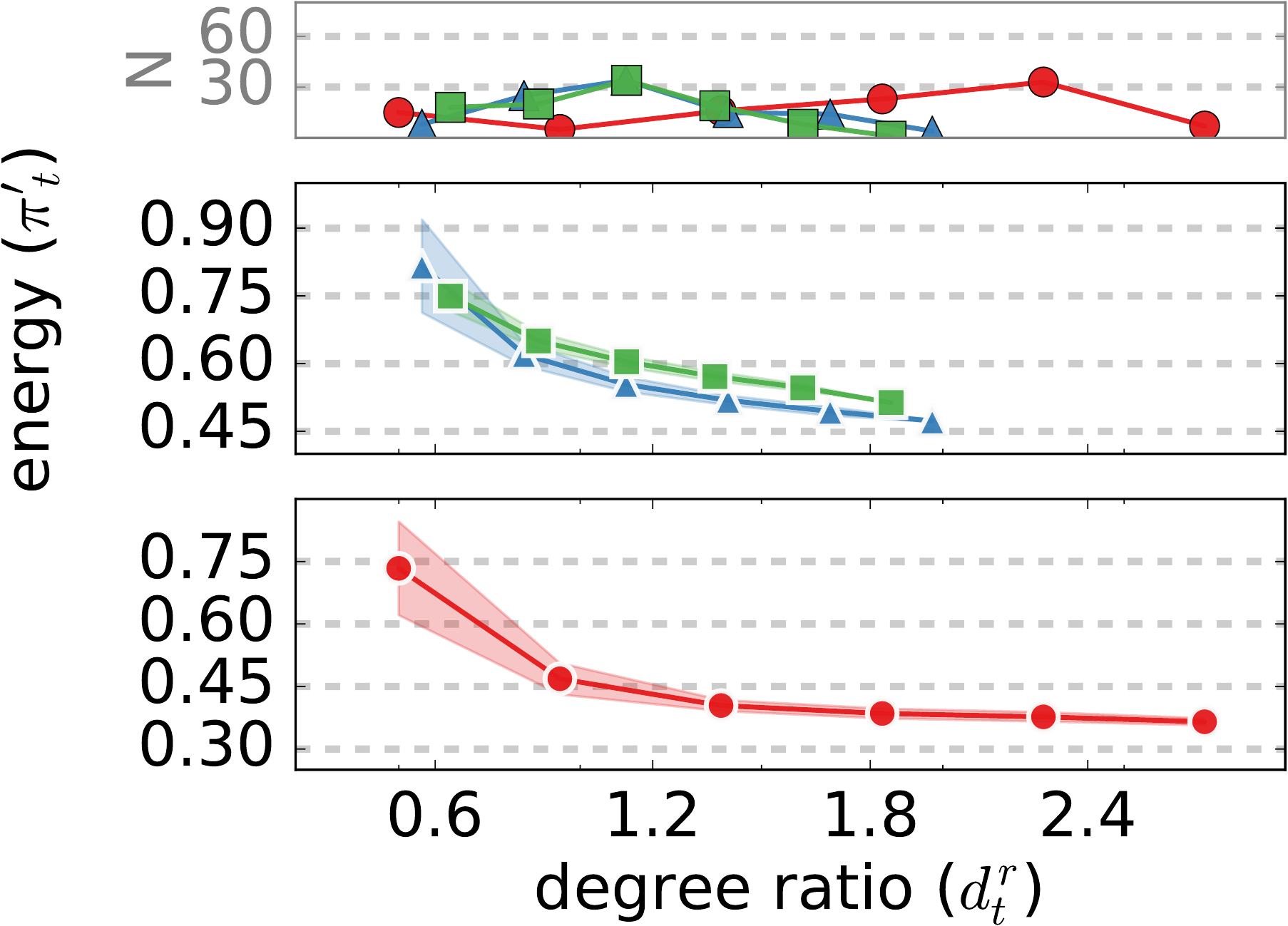}}\hfill
\subfloat[\term{DEM}\xspace]{\includegraphics[width=.3\linewidth]{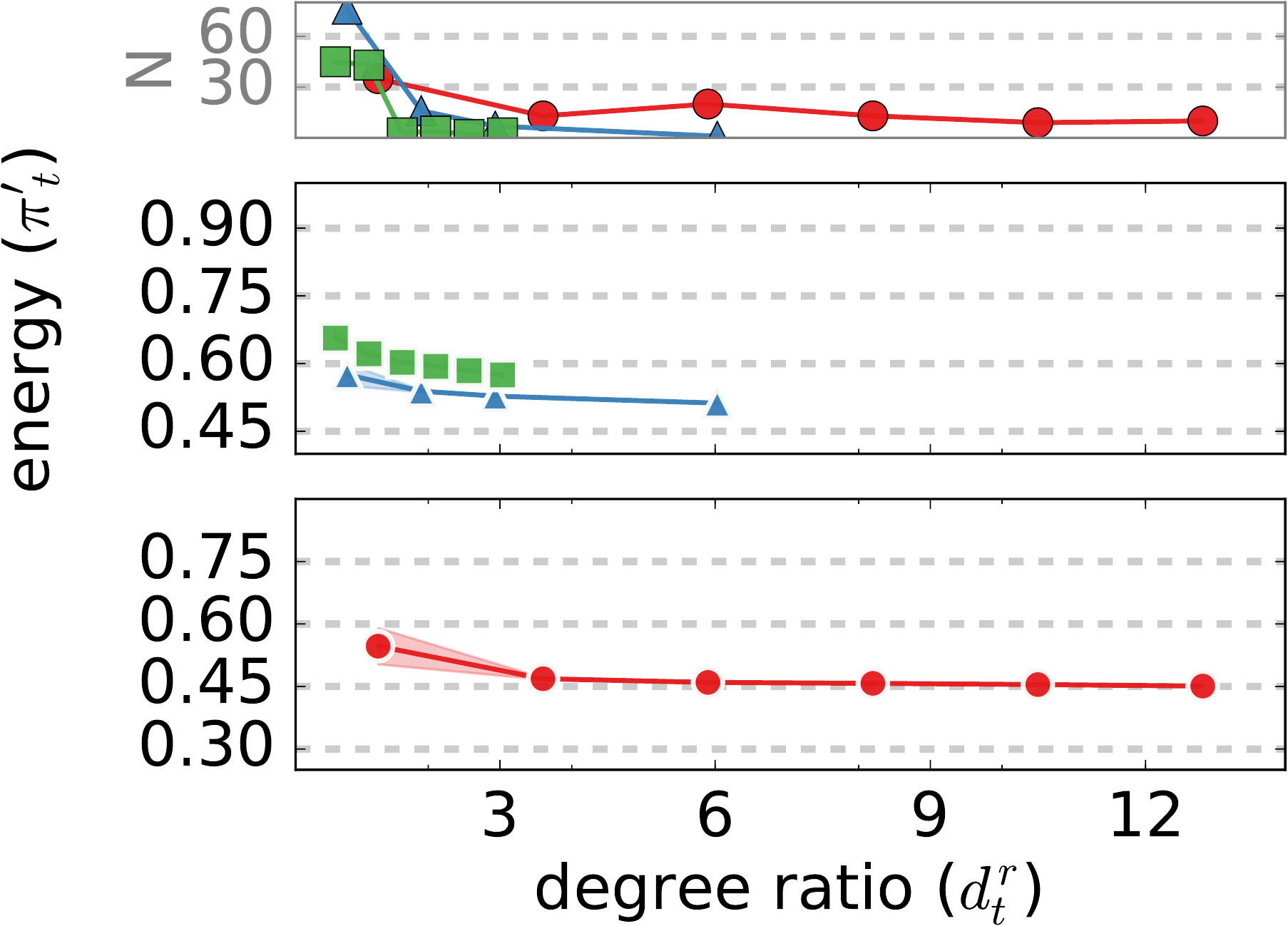}}
\hspace*{\fill}
	\fi
	\vspace*{-.8\baselineskip}
	\caption{
		\textbf{Influence of Target Nodes Degree Ratio onto the Saturation of their Energy.} The plots depict the due to link insertion achieved energy of target nodes $pi'_t$ as the function of their degree ratio. Each line depicts the results for a given $\phi$. For increased readability we group data points into six equally sized bins according to their degree ratio ($x$-axes). Values on the $y$-axes represent the averages of the data points falling into the corresponding bin. \textbf{Top row.} In the top row we show the distribution of the target nodes degree ratios (in bins; $x$-axes)---the $y$-axes denote the number of data points (N) falling into each bin. \textbf{Middle row.} Here, we depict the results for medium and large fractions of target nodes. \textbf{Bottom row.} For readability we depict small fractions of target nodes separately. Over all datasets and all $\phi$ we consistently observe a negative correlation between degree ratio and energy $\pi'_t$. This means that with an increasing ratio---an increasing out-degree and a decreasing in-degree---the drain of energy increases and this leads to the saturation of the energy of target nodes.
	}
	\label{fig:stat_prob:struc_prop}
	\vspace*{-1.2\baselineskip}
\end{figure*}

Figure~\ref{fig:saturation} depicts the effects of link modifications in our datasets with increasing values of bias strength $b$ and varying fractions of target nodes $\phi$ ($0.01$, $0.1$ and $0.2$). 

In the case of click bias we observe the following situation. For small values of $b$ the energy of target nodes $\pi'_t$  increases very quickly (navigational boost phase, which we analyze in more detail in Section \ref{sec:results:navboost})---this energy saturates for larger values of $b$ (i.e., $b > 35$). This holds for larger $\phi$ values ($0.1$ and $0.2$), whereas for a smaller $\phi$, for example $\phi=0.01$, the initial growth as well as the saturation are significantly slower and lower respectively. Further, for higher $\phi$ ($0.1$ and $0.2$) $\pi'_t$ saturates at an almost identical and very high level (>0.8)---if the click bias is strong enough we can increase the energy of any fraction of target nodes larger than $\phi>0.1$.

An interesting question in this respect is the height of the energy saturation level. Theoretically, this level is close to $1.0$ but as Figure~\ref{fig:saturation} shows, in empirical networks this level can not be fully reached. Essentially, due to the directed nature of the network, the target nodes out-component (i.e., the nodes with incoming links from  target nodes) will always act as a drain that will take some energy from the target nodes. That amount depends on the size of the out-component as well as its connectivity with other parts of the network---in particular the existence of back-links towards target nodes. This situation is depicted in our toy example Figure~\ref{fig:edu} in the middle row. Node $4$, which has an incoming link from node $1$, profits from an induced click bias towards node $1$ (cf. original $\pi_4=0.27$ and modified $\pi'_4=0.29$). Thus, although $\pi'_1$ increases with increasing bias strength, node $1$ would never reach energy values close to $1.0$ because node $4$ attracts a certain amount energy to itself.

In the case of the link insertion strategy the results are more diverse (cf. Figure~\ref{fig:stat_prob:LI}). For \term{DEM}\xspace dataset we observe a quick saturation for all values of $\phi$. Differently from the click bias the saturation level is significantly lower for this dataset (i.e., $0.6$). For the \term{ORF}\xspace dataset we do not observe saturation but a monotonous increase in the energy of target nodes for increasing values of $\phi$. Finally, for the \term{W4S}\xspace dataset and larger $\phi$ ($0.1$ and $0.2$) we can observe saturation at levels higher than $0.9$. 

As previously, the size of the out-component of the target nodes, combined with the size of their in-component (i.e., the source nodes which point towards target nodes), as well as the ratio of these two quantities provide a possible explanation for this behavior. Basically, we can calculate the average number of newly inserted links as $l(b)=\overline{d} \cdot n \cdot \phi \cdot b$, where $\overline{d}$ is the average degree (i.e., in a directed network average degree $\overline{d}$ corresponds to both the average in-degree as well as average out-degree) and $n$, $\phi$, $b$ are as before. Thus, in the networks with a higher average degree we insert more new links. For smaller values of bias strength (blueish region in Figure~\ref{fig:stat_prob:LI}) these new links lead to a navigational boost, resulting in a quick increase in the energy $\pi'_t$ of target nodes. The navigational boost is higher in networks with a higher average degree---we observe the highest increase in $\pi'_t$ in \term{DEM}\xspace with $\overline{d}=49.22$, the second highest in \term{ORF}\xspace with $\overline{d}=30.8$, and the lowest in \term{W4S}\xspace with $\overline{d}=27.6$. As mentioned before, in Section \ref{sec:results:navboost} we analyze this navigational boost in more detail. However, for larger values of bias strength (reddish region in Figure~\ref{fig:stat_prob:LI}) the effects of the drain due to the larger size of the out-component become visible---the networks with a higher increase for smaller bias strengths lose their energy now more quickly. Thus, the ordering of the saturation levels for higher bias strengths is reversed to the navigational boost in energy for lower bias strengths, resulting in \term{W4S}\xspace to now have the highest saturation level, followed by \term{ORF}\xspace and then by \term{DEM}\xspace.

To confirm our intuition about the saturation for the link insertion strategy we performed the following analysis. First, we calculated some structural properties for the target nodes. In particular, based on the insights of Ding et al.~\cite{ding2002pagerank, ding2004link}, we define the \textit{in-degree} of target nodes as the sum of the weights of links pointing towards target nodes $d_t^- = \sum_{ij} \left(\text{ diag}(\bm{t}) \cdot \bm{W} \right)_{ij}$. The \textit{out-degree} of target nodes is the sum of the weights of outgoing links of target nodes $d_t^+ = \sum_{ij} \left(\bm{W} \cdot \text{ diag}(\bm{t})\right)_{ij}$. Finally, the \textit{degree ratio} of target nodes is a ratio between the previous two measurements (i.e., $d_t^r = d_t^+/d_t^-$). Although, it has been shown that properties, such as the simple count of in-links of a node, are bad approximations for PageRank on a large scale~\cite{pandurangan2002using}, they proved to be a good indicator for the random surfer behavior on our datasets.

In our experiments, \term{DEM}\xspace has on average by one order of magnitude higher both target node in-degree and out-degree than the other two datasets. This explains a quick increase of $\pi'_t$ for smaller bias strengths. However, degree ratio is typically 
larger in \term{DEM}\xspace target nodes than in \term{ORF}\xspace or \term{W4S}\xspace target nodes and this explains a higher drain of energy and a lower saturation level in the \term{DEM}\xspace dataset (cf. Figure \ref{fig:stat_prob:struc_prop}).

\findingbox{For larger fractions $\phi$ of target nodes their energy $\pi'_t$ achieved through a click bias quickly saturates across all datasets at very high levels ($>0.8$). Boost and saturation of the energy is significantly slower for smaller fractions $\phi$. The saturation level is determined by the out-degree of the target nodes and reciprocity of outgoing links from the target nodes. For link insertion saturation existence, speed, and levels vary between datasets and $\phi$ values. The average degree of the original networks as well as the ratio between out-degree and in-degree of target nodes significantly influences those effects.}

\noindent\textbf{Implications.} In case of medium ($\phi=0.1$) and large ($\phi=0.2$) fractions of target nodes we reach high saturation levels with both link modification methods even with small bias strengths. For example, if we would like to increase visibility of a large category in, for example Wikipedia, we can achieve this by either slightly increasing the font size of the links towards the articles of that category or by simple creating some new links towards those articles. Click bias reaches very high visibility levels consistently across several different datasets, whereas link insertion is dependent on the network structure---in datasets with a smaller average number of links we can achieve larger changes. This follows our intuition---in a network with a smaller number of links, each new link affects the network more significantly. However, to match the effects of the click bias we need to insert a very large amount of new links. On the other hand, in case of small ($\phi=0.01$) fractions of target nodes, we can achieve larger changes by using link insertion---we are able to reach higher saturation levels more consistently and more quickly regardless of the dataset. Again, we can explain this intuitively---small fractions of target nodes have, on average, only few links pointing towards them. Hence, inserting a new link from a top webpage\xspace achieves larger changes than highlighting an existing (and probably negligible) link.

\subsection{Navigational Boost}
\label{sec:results:navboost}

\begin{figure*}[ht!]
	\renewcommand{\FigBiasStrength}{0005}
	\captionsetup[subfigure]{justification=centering}
	\captionsetup[subfloat]{farskip=2pt,captionskip=1pt}
	\centering
	\renewcommand{\figwidth}{.3\linewidth}
	\includegraphics[width=.4\linewidth]{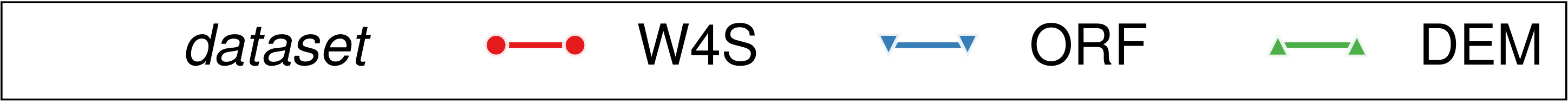} \\
	\hspace*{\fill}
	\subfloat[Click Bias]{\includegraphics[width=.3\linewidth]{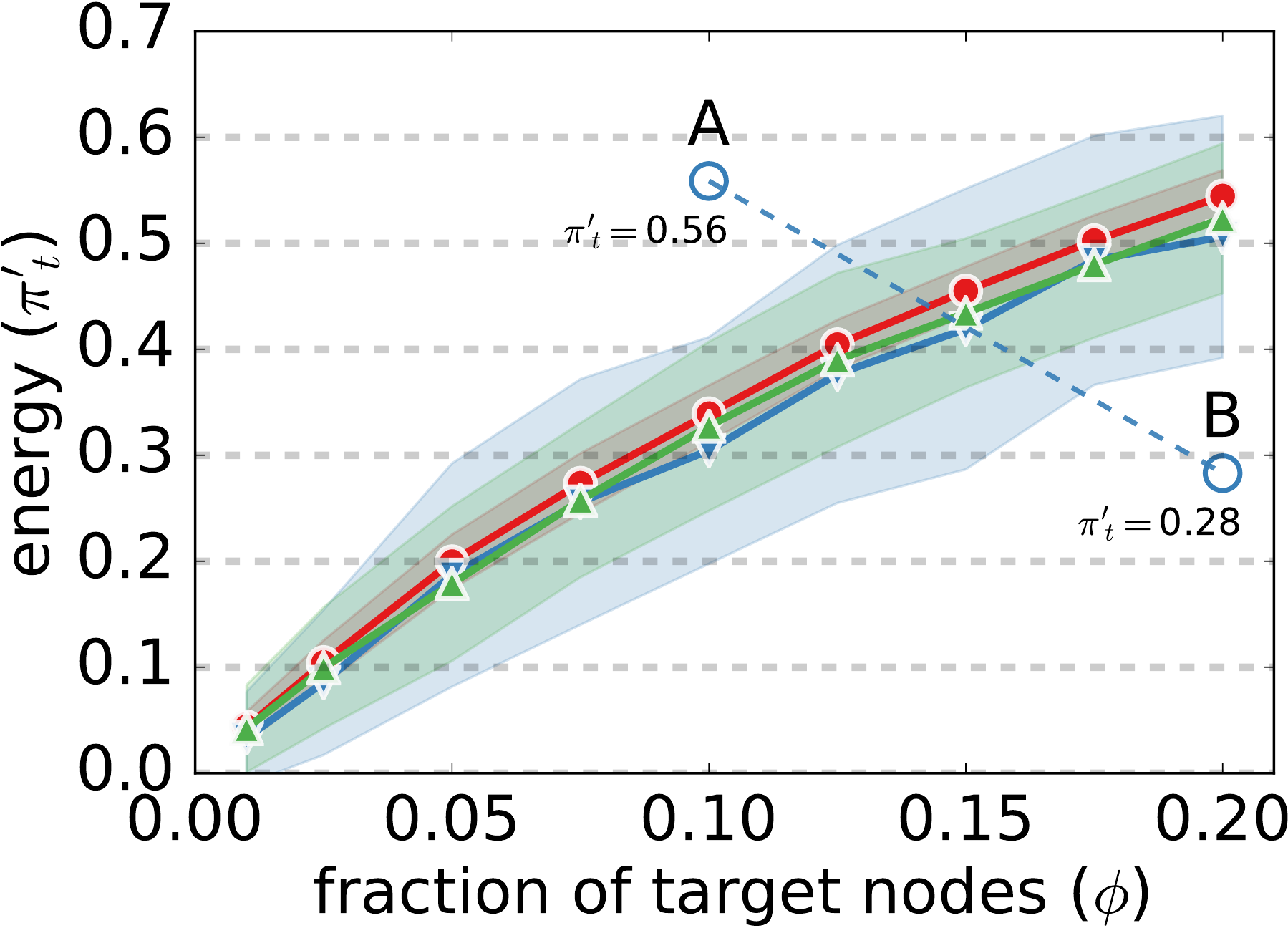}\label{fig:stat_prob:bias}}\hfill
	\subfloat[Link Insertion]{\includegraphics[width=.3\linewidth]{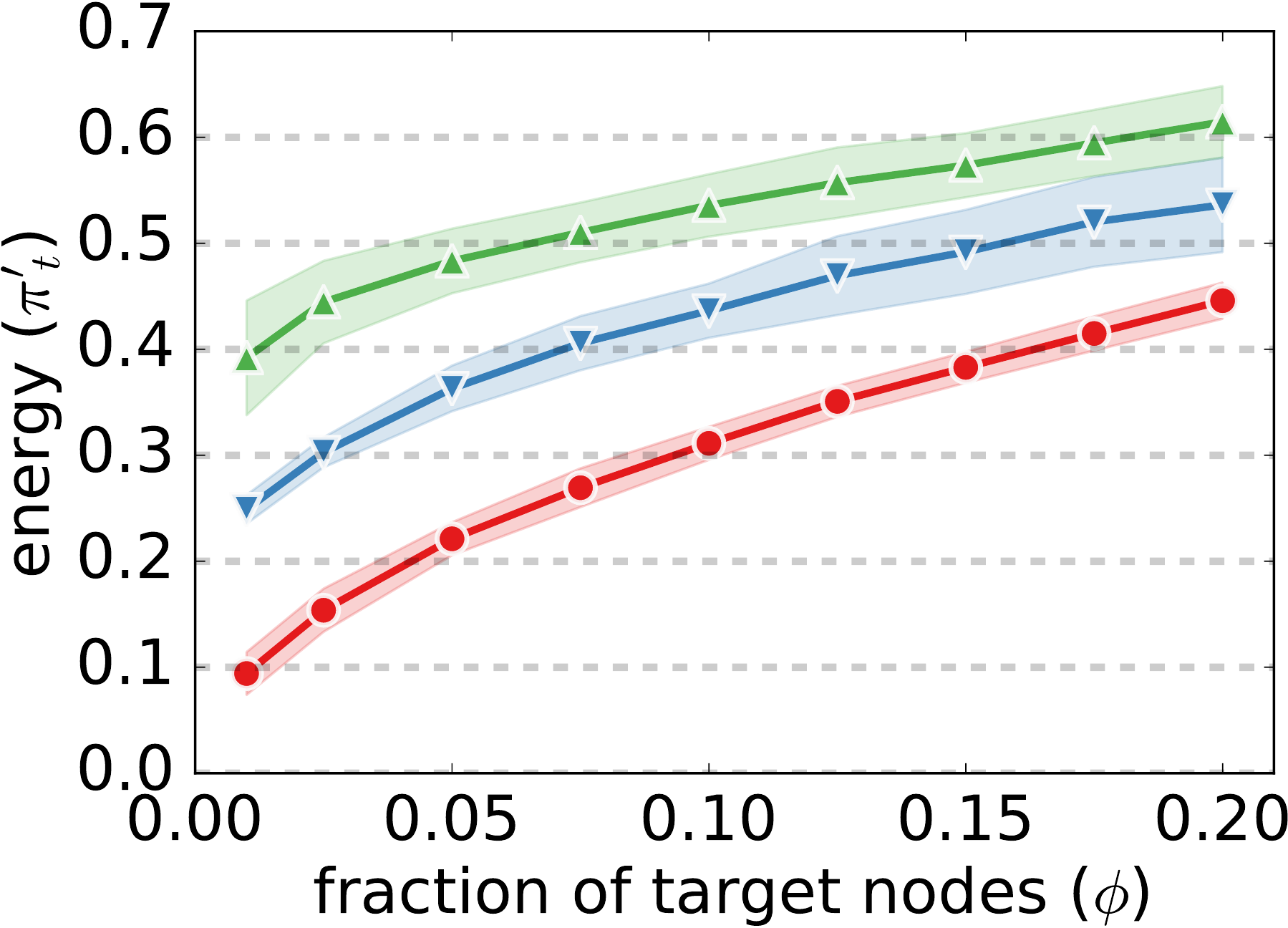}\label{fig:stat_prob:li}}\hfill
	\subfloat[Lorenz Curve]{\includegraphics[width=.3\linewidth]{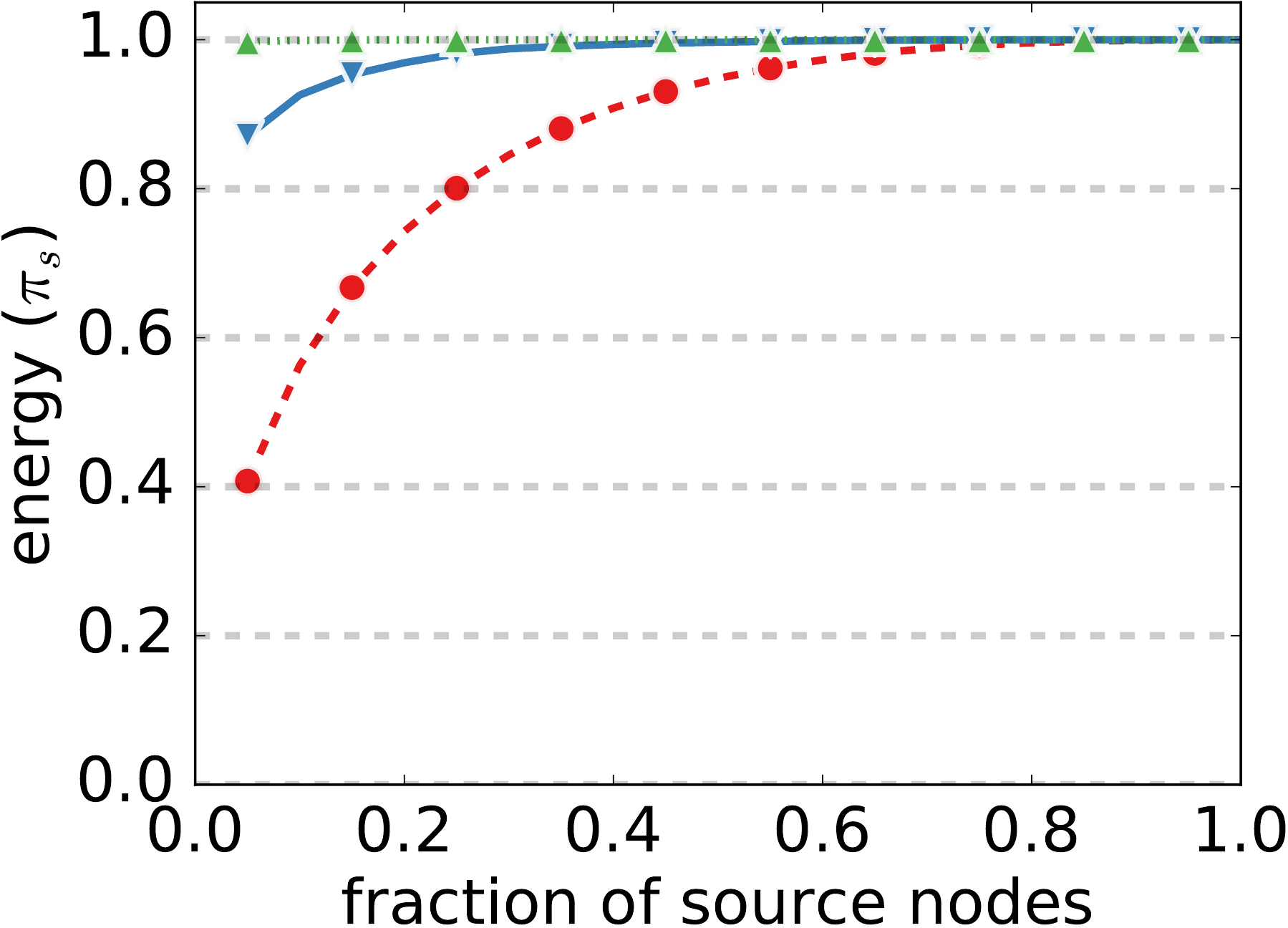}\label{fig:stat_prob:ccdf}} \hfill
	\hspace*{\fill}
	\vspace*{-.8\baselineskip}
	\caption{\textbf{Navigational Boost.} Left center figures depict the energy of target nodes after modifying the network through either inducing a click bias or link insertion respectively. The $x$-axes correspond to the fraction $\phi$ of target nodes, whereas on the $y$-axes we denote the energy of target nodes $\pi'_t$. Each line represents the average of $100$ samples for each $\phi$ of one dataset. The areas filled with the same color denote the corresponding standard deviations. \textbf{Left.} Inducing a click bias is robust across datasets. However, the variability within values of $\phi$ is rather high. The high variance is caused by the presence or absence of one or more nodes with high original energy in the target nodes. Thus, in cases where such nodes are present in target nodes (depicted as point A in the plot) even smaller fractions of target nodes are able to outperform larger fractions of target nodes without such a top node (depicted as point B in the plot). \textbf{Center.} On the contrary, the performance of link insertion varies over datasets but is stable across various $\phi$ values, which is signified by the low standard deviation suggests over $\phi$. \textbf{Right.} We plot the Lorenz curve of the datasets' original stationary distributions. We can observe that for different datasets these distributions are differently skewed. In particular, in \term{DEM}\xspace the energy of just a few nodes is close to $1$, whereas in \term{ORF}\xspace and \term{W4S}\xspace we need far more nodes to reach the same level (i.e., $~0.4$ and $~0.7$ respectively). This explains why we can achieve the highest effect with link insertion in \term{DEM}\xspace, followed by \term{ORF}\xspace and \term{W4S}\xspace. Thus, the performance of link insertion depends on the initial stationary distribution of the network, whereas click biases are robust across datasets. Moreover, for smaller fractions of target nodes link insertion constantly outperforms click biases.}
	\label{fig:stat_prob}
	\vspace*{-1.2\baselineskip}
\end{figure*}

The blueish region from Figure~\ref{fig:stat_prob:LI} corresponds to smaller and more realistic bias strengths. In practice, increasing the visibility of a link (for example, by repositioning or increasing the font size) by more than a factor of $15$, meaning that it would receive $15$ times more clicks than before, seems quite unrealistic. In particular, users position bias is estimated to be lower than $3.5$~\cite{lerman_pos_bias, hoggDisentangling}.
Hence, we focus on bias strengths ranging from $2$ to $15$ (the blueish region in Figure~\ref{fig:stat_prob:LI}) where we can observe a phase of quick increase in the energy of target nodes. We call this phase \textit{navigational boost} phase. The results for all bias strengths from $2$ to $15$ are quite similar and therefore we report only the results for bias strength $b=5$.

For click bias we observe a robust performance across datasets, see Figure \ref{fig:stat_prob}. The energy of target nodes $\pi'_t$ increases almost linearly with the fraction $\phi$ of target nodes. However, at higher $\phi$ (i.e., $0.15 \le \phi \le 0.2$) the linear trend tends to flatten. This is due to a transition to the stationary phase (cf. Section \ref{sec:results:saturation}). Further, we observe a rather high variance of $\pi'_t$ over $\phi$ and different sets of target nodes. For example, we measure the following average standard deviations over $\phi$: \term{W4S}\xspace $=0.023$, \term{ORF}\xspace $=0.103$ and \term{DEM}\xspace $=0.068$. This high variance can be attributed to situations in which smaller fractions of target nodes are often able to outperform larger ones. We depict one such extreme situation of two outlier samples marked as \textit{A} and \textit{B} in Figure \ref{fig:stat_prob}. Target nodes depicted with \textit{A} with $\phi=0.1$ reach an energy that is almost twice as high as those of the target nodes depicted with \textit{B} with $\phi=0.2$. 

One potential explanation for these observations is that if the energy of target nodes of the unmodified network is already quite high, that is, the target nodes include one or more nodes with a substantial energy, then the click bias acts as an \textit{amplifier} further magnifying the energy of target nodes. On the other hand, target nodes with a small unmodified energy receive indeed the amplifying effect but are never able to reach the same (high) levels of the modified energy. Therefore, it is possible for smaller fractions of target nodes with one or more nodes with high starting energy to outperform larger fractions of target without such nodes.
This can be further attributed to the target nodes structural properties, such as out-degree, in-degree and degree ratio, which we introduced in the previous sections. Basically, starting energy positively correlates with in-degree of target nodes, and therefore we can expect that the click bias is able to amplify target nodes with a higher in-degree more than the target nodes with a lower in-degree. In particular, to confirm this finding we conducted a similar correlation experiment as depicted in Figure~\ref{fig:stat_prob:struc_prop}, but used a combination of the target nodes in-degree and energy achieved due to a click bias. However, due to limitations in space, we do not report the experimental details here. %

\findingbox{The fraction $\phi$ of target nodes does not have a decisive effect on navigational boost. Often, smaller $\phi$ exhibit larger effect sizes. Click bias acts as an \textit{amplifier} that only magnifies what is already present in the target nodes.}

In the case of link insertion, navigational boost appears to be highly dataset dependent (see Figure~\ref{fig:stat_prob:li}). However, the variance of each dataset individually is very low with average standard deviations of $0.017$ for \term{W4S}\xspace, $0.029$ for \term{ORF}\xspace and $0.034$ for \term{DEM}\xspace. Across all datasets we can observe a quick increase in the energy of target nodes with an increasing fraction of target nodes, which then experience a transition towards a stable saturation phase. 

To explain the difference in performance between different datasets we have plotted the Lorenz curves of the stationary distributions of our datasets (see Figure~\ref{fig:stat_prob:ccdf}). We see that for \term{W4S}\xspace, a very small fraction of top nodes ($0.01$) only possesses $0.4$ of energy. Diversely, for \term{ORF}\xspace and \term{DEM}\xspace the same fraction of top nodes already possesses energy higher than $0.85$. As the out-component of a specific set of nodes acts as a drain for the energy of source nodes, connecting source nodes with high energy to target nodes leads to a flow of energy from those source nodes towards target nodes. Thus, the initial energy of source nodes plays a crucial role in this process. Through link insertion from top source nodes towards target nodes we attach the target nodes as drains to such top nodes. Consequently, target nodes receive a huge amount of energy and experience a large navigational boost (i.e., \term{ORF}\xspace and \term{DEM}\xspace). In other words, we can say that link insertion induces \textit{diffusion} of the energy of top nodes towards target nodes. Given the average degree of the network and the fraction of source nodes (which increases with the fraction $\phi$ of target nodes), we can use the Lorenz curves to approximately predict the point where the performance across datasets becomes similar. For example, the Lorenz curves of \term{DEM}\xspace and \term{ORF}\xspace meet around a fraction of $0.4$ of source nodes and we can expect that the performance of those two datasets will become similar for all fractions of source nodes larger than $0.4$. In the case of \term{W4S}\xspace, we need a larger fraction of source nodes ($0.7$) to reach a similar behavior (cf. Figure~\ref{fig:stat_prob:ccdf}).

Comparing link insertion with click bias we find that the former outperforms the latter for smaller fractions $\phi$ of target nodes. For example, in the \term{DEM}\xspace dataset, link insertion reaches four times higher energy values for the target nodes with $\phi = 0.01$. However, for higher values of $\phi$ the click bias exhibits a similar performance as link insertion. Further, in the case of the \term{W4S}\xspace dataset, the click bias even outperforms link insertion (see \term{W4S}\xspace in Figure~\ref{fig:stat_prob:bias} and Figure~\ref{fig:stat_prob:li} at $\phi=0.2$). 

\findingbox{The performance of link insertion varies across the datasets and depends on the skewness of the initial stationary distribution in a dataset. Inserting links from other important webpages\xspace towards a given set of webpages\xspace results in a higher navigational boost than with the click bias. This is due to the induced \textit{diffusion} of the energy from top nodes towards target nodes.}

\noindent\textbf{Implications.} If it is possible to insert new links on a website\xspace (especially if the fraction of target nodes is small) we should prefer the link insertion over the click bias. 
However, creation and insertion of such links may be problematic in practice. For example, on Wikipedia it may be difficult and semantically unjustified to insert new links to completely unrelated articles since this may have opposite and contrasting effects on the navigational behavior of users, such as confusion and dissatisfaction. In those cases we may rather choose to increase the transition probability of an already existing link by, for example, highlighting that link (i.e., using CSS\footnote{cascading stylesheets}) or repositioning it to the webpage\xspace's top area. In some other scenarios (i.e., birthdays of famous inventors) implementing a banner which contains links towards a given set of webpages\xspace may be an easy way to insert thousands of new links instantly. In those cases, such user interface modifications may prove to have higher lasting effects on the stationary probability than, for example, highlighting links.

\subsection{Influence Potential}
\label{sec:results:mod_pot}

\begin{figure}[t!]
	\renewcommand{\FigBiasStrength}{0005}
	\captionsetup[subfigure]{justification=centering}
	\captionsetup[subfloat]{farskip=2pt,captionskip=1pt}
	\centering
	\renewcommand{\figwidth}{.48\linewidth}
	
	\includegraphics[width=.7\linewidth]{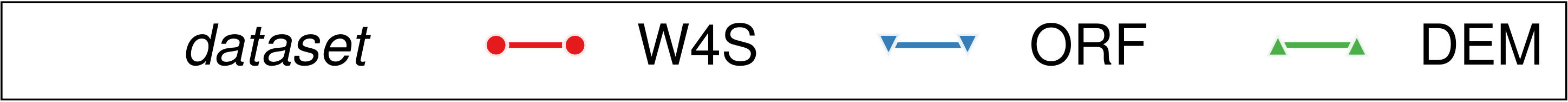}
	
	\hspace*{\fill}
	\subfloat[Click Bias]{\includegraphics[width=.48\linewidth]{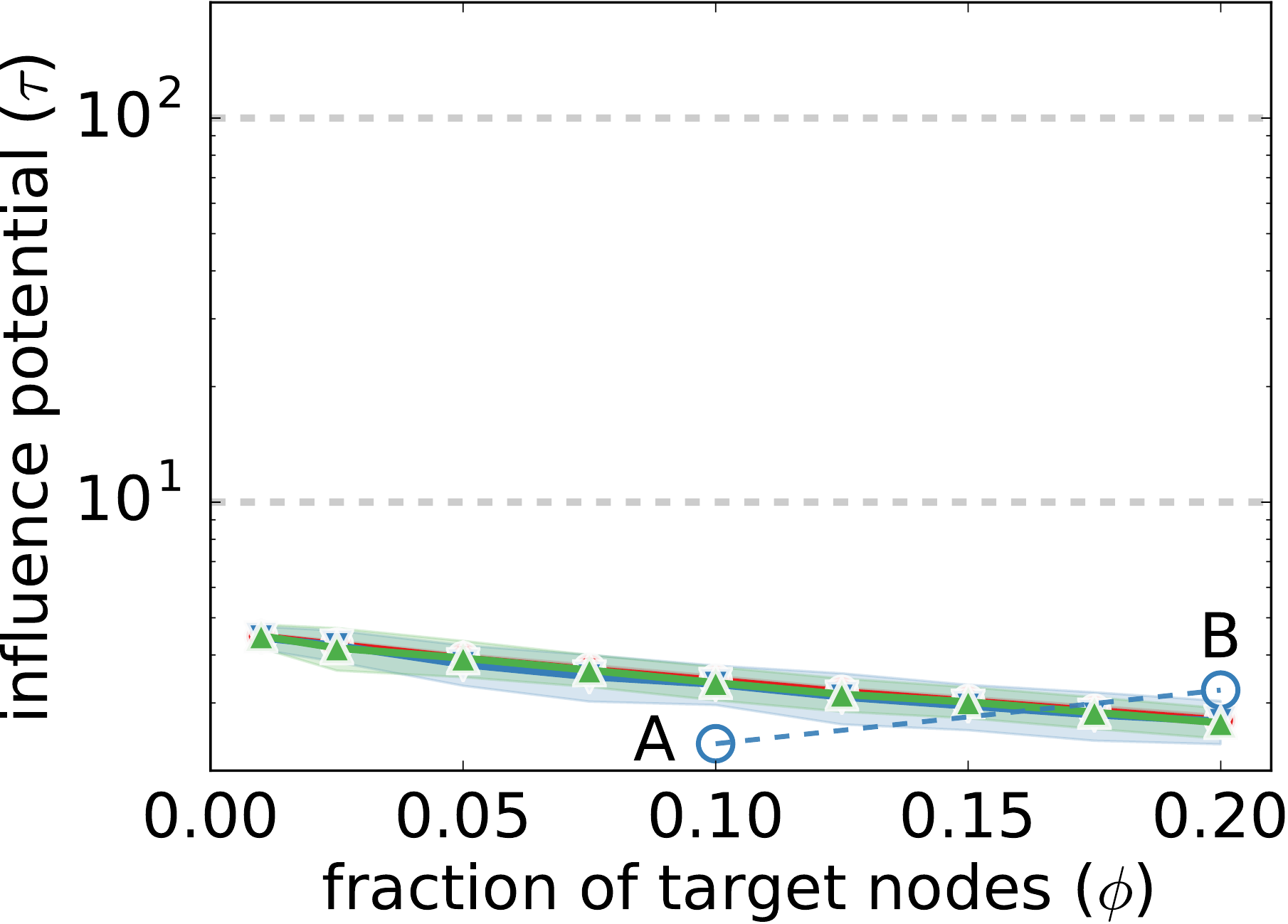}\label{fig:mod_pot:bias}}\hfill
	\subfloat[Link Inseration]{\includegraphics[width=.48\linewidth]{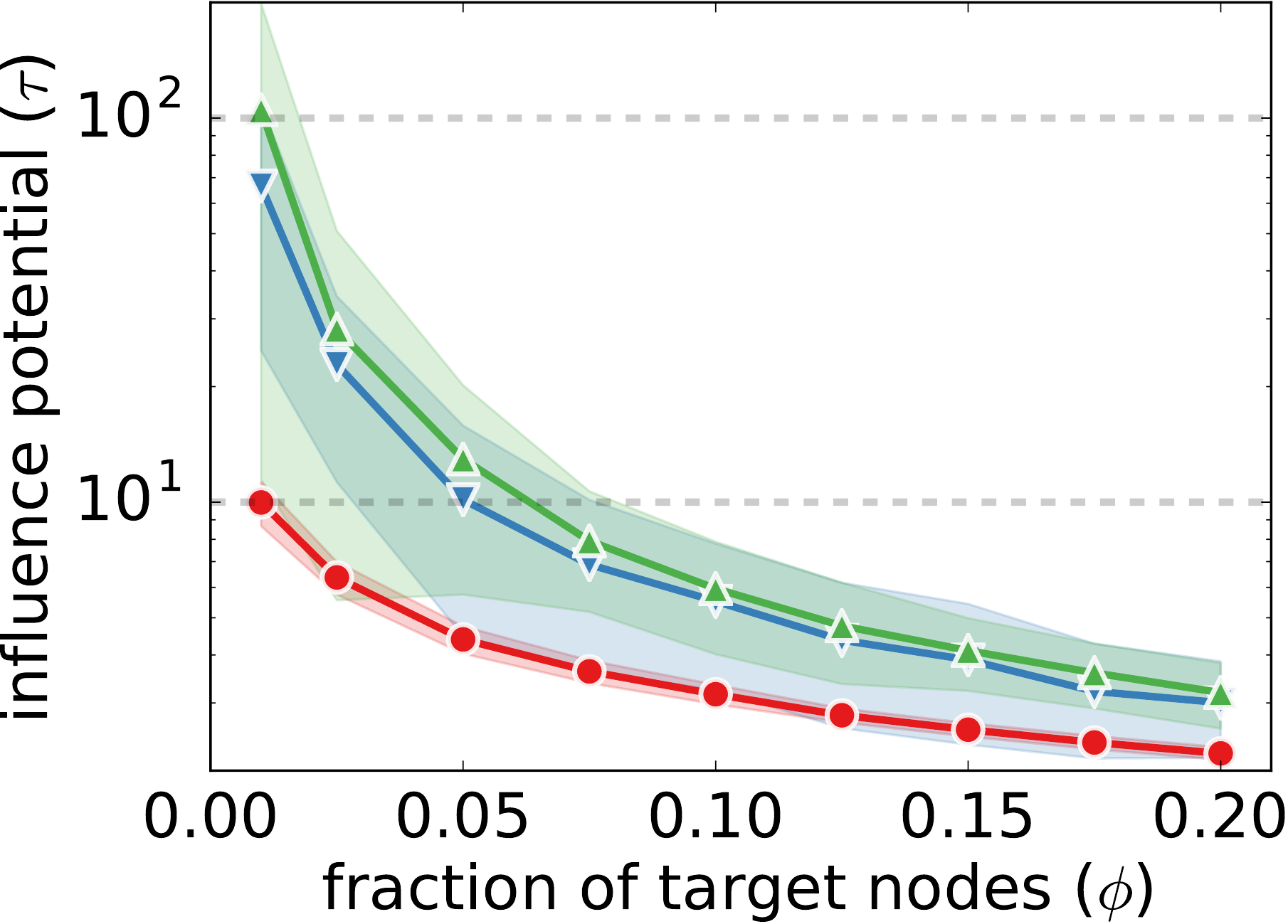}\label{fig:mod_pot:li}}\hfill
	\vspace*{-.8\baselineskip}
	\caption{\textbf{Influence Potential.} The figure depicts the relative increase in the energy of target nodes $\tau$ with a fixed bias strength of $b=5$ over different $\phi$ and datasets. \textbf{Left.} Inducing a \textit{click bias} performs robustly over all datasets. Influence potential correlates negatively with the fraction of target nodes, that is, the relative increase in energy is higher for small fractions of target nodes than for large fractions. \textbf{Right.} With link insertion, we find a significant variance in performance across the three different datasets. This confirms our findings from the previous section---the skewness of the original stationary distribution determines the effectiveness of the link insertion strategy in a dataset. Similarly to the click bias, the influence potential decays with an increasing fraction of target nodes.}
	\label{fig:mod_pot}
	\vspace*{-1.2\baselineskip}
\end{figure}

\begin{figure*}[ht!]
	\captionsetup[subfigure]{justification=centering}
	\captionsetup[subfloat]{farskip=2pt,captionskip=1pt}
	\centering
	\renewcommand{\figwidth}{.3\linewidth}
	
	\renewcommand{\figrow}[2]{
		\hspace*{\fill}
		\subfloat[$\phi = 0.01$]{\includegraphics[width=.3\linewidth]{#1_ss_0_010_mean}\label{fig:combinations:001}}\hfill
		\subfloat[$\phi = 0.1$]{\includegraphics[width=.3\linewidth]{#1_ss_0_100_mean}\label{fig:combinations:01}}\hfill
		\subfloat[$\phi = 0.2$]{\includegraphics[width=.3\linewidth]{#1_ss_0_200_mean}\label{fig:combinations:02}}
		\hspace*{\fill}
	}
	
	\includegraphics[width=.5\linewidth]{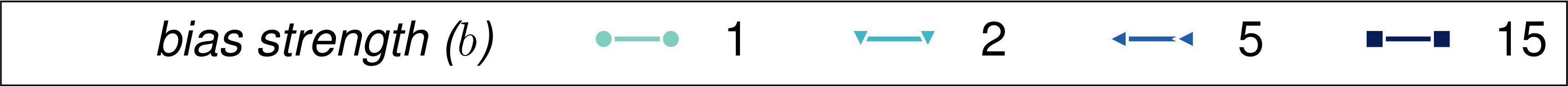}
	
\hspace*{\fill}
\subfloat[$\phi = 0.01$]{\includegraphics[width=.3\linewidth]{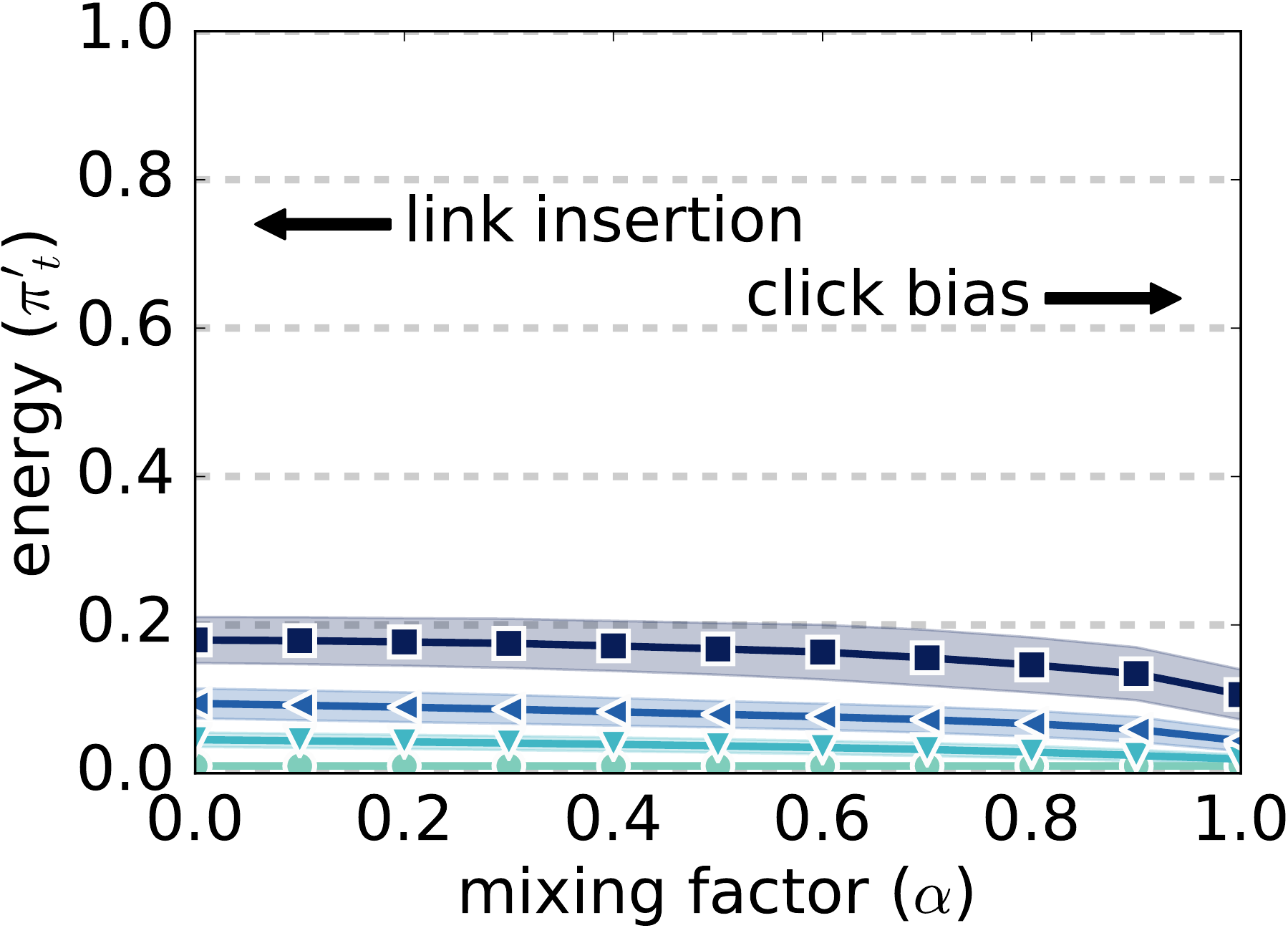}\label{fig:combinations:001}}\hfill
\subfloat[$\phi = 0.1$]{\includegraphics[width=.3\linewidth]{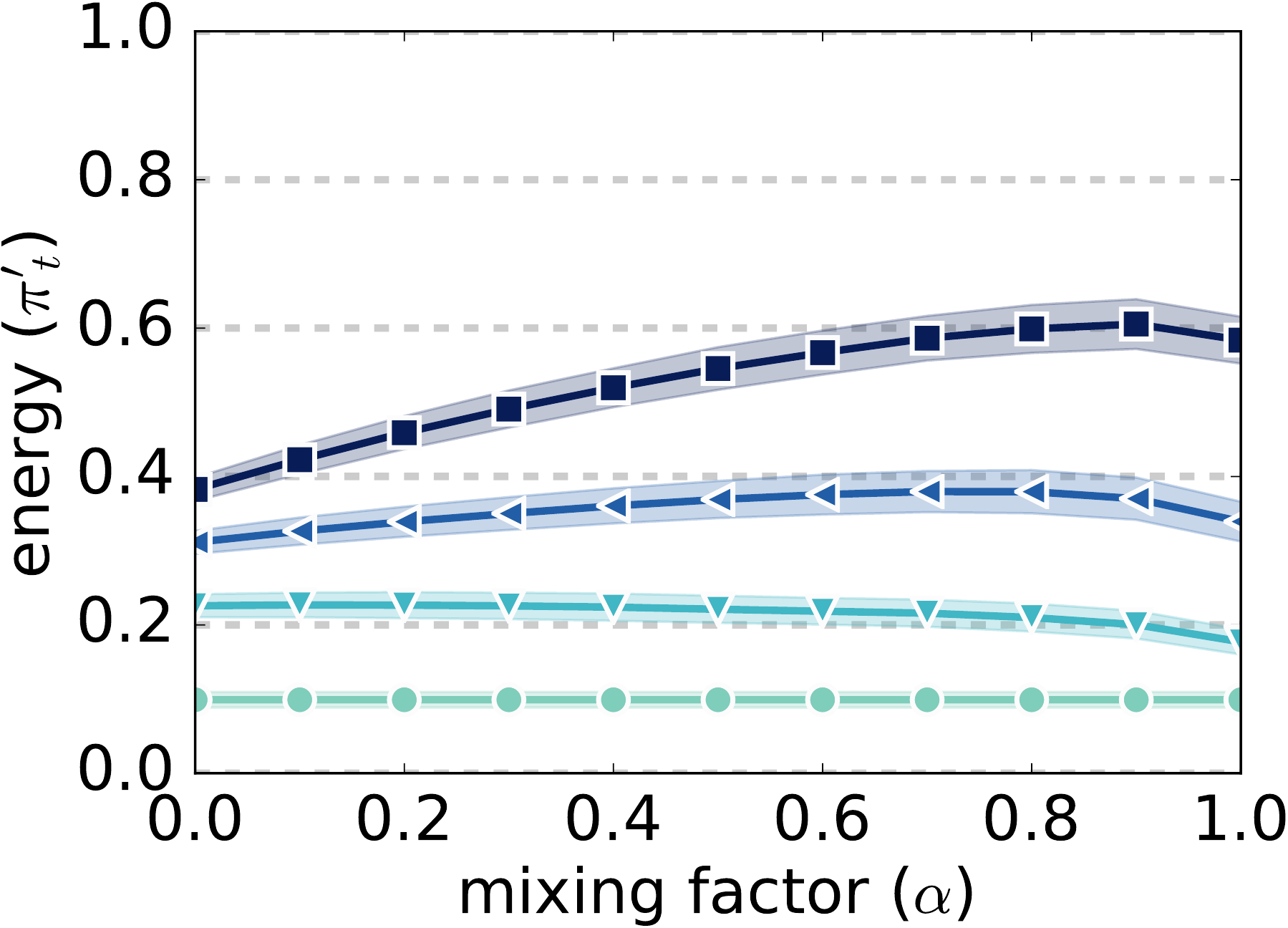}\label{fig:combinations:01}}\hfill
\subfloat[$\phi = 0.2$]{\includegraphics[width=.3\linewidth]{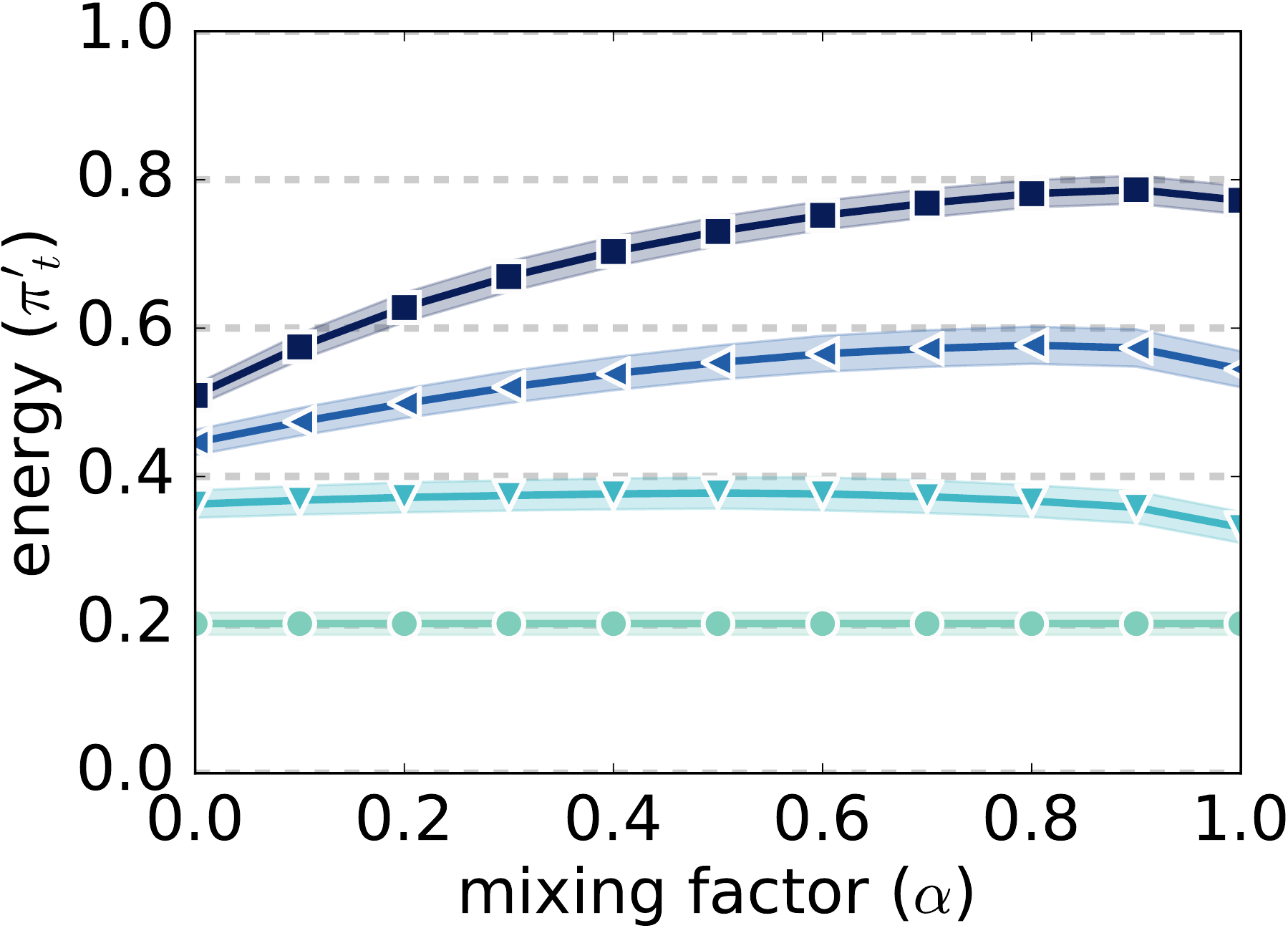}\label{fig:combinations:02}}
\hspace*{\fill}
	
	\vspace*{-.8\baselineskip}
	\caption{
		\textbf{Combinations of Link Modification Strategies.} The plots depict average results of $100$ sets of target nodes of \term{W4S}\xspace for three $\phi$: $0.01$ small on the left, $0.1$ medium in the middle, and $0.2$ large on the right. On the $x$-axes we denote $\alpha$, which defines the combination of the two link modification strategies, whereas on the $y$-axes we denote the energy of target nodes. We see that for smaller values of $\phi$ $\alpha=0$ ($100\%$ link insertion) outperforms all the others over all used bias strengths. However, for medium and large $\phi$ values with higher bias strengths this sweet spot shifts towards higher combinations ($\alpha=0.7$). In the other two datasets we can observe similar results.
	}
	\label{fig:combinations}
	\vspace*{-1.2\baselineskip}
\end{figure*}

Figure~\ref{fig:mod_pot} depicts the effects of link modifications strategies on the relative increase of the energy of target nodes (i.e., influence potential). Again, we concentrated in this experiment on realistic settings for the bias strength from the interval $[2, 15]$. Since we got comparable results over that complete interval we present only the results for bias strength $b=5$.

The performance of the click bias is robust across datasets and different $\phi$ with a low variance in both dimensions (cf. Figure \ref{fig:mod_pot:bias}).
We observe a negative correlation between influence potential and fraction $\phi$ of target nodes, meaning the smaller fractions of target nodes profit more from an induced click bias than larger fractions. Our calculations of the influence potential confirm once more the results from the previous section, in which smaller fractions with top energy nodes are able to outperform larger fractions of target nodes without top nodes. We once more depict two such examples from Figure~\ref{fig:stat_prob:bias}. Target nodes depicted by \textit{A} with $\phi=0.1$ reach an energy that is almost twice as high as those depicted by \textit{B} with $\phi=0.2$. However, nodes \textit{A} start with a larger initial energy and nodes \textit{B} with a smaller one. Therefore, in relative terms nodes \textit{B} have a higher influence potential than nodes \textit{A} (cf. Figure \ref{fig:mod_pot:bias}).

Performance of link insertion is again strongly dependent of the dataset. However, similarly to the click bias we observe over all datasets that smaller fractions of target nodes profit significantly more from the link insertion than the larger ones. For example, in \term{DEM}\xspace dataset for $\phi=0.01$ we measure an average influence potential of more than $100$, whereas for $\phi=0.2$ influence potential is less than $4$ (cf. Figure~\ref{fig:mod_pot:li}). A similar decay, although not as pronounced as in \term{DEM}\xspace can be seen in the other two datasets. Similarly to the navigational boost this high influence potential of smaller fractions of target nodes in the case of link insertion can be explained through the skewness of the initial stationary distributions (cf. Figure~\ref{fig:stat_prob:ccdf}).

As previously, we investigated more closely the relation between influence potential of small fractions of target nodes and their structural properties such as in-degree, out-degree and degree ratio. Target nodes with a high degree ratio (i.e., a small in-degree, a large out-degree or both) have the largest influence potential. Intuitively, such target nodes start with a very small initial energy and therefore can achieve a significant relative increase. On contrary, in absolute terms such target nodes keep a rather small energy even after the modification, whereas target nodes with a large initial energy (a low degree ratio) are experiencing a significant navigational boost in absolute terms but possess relatively low influence potential.

\findingbox{The influence potential of small fractions of target nodes is very high regardless of the link modification strategy. For click bias the influence potential is limited by the bias strength, whereas for link insertion we do not observe such a limit and influence potential can become as high as $100$. With increasing fraction of target nodes the influence potential decays drastically.}

\noindent\textbf{Implications.} As previously, if possible we should prefer link insertion over click bias in cases where we are interested in utilizing the influence potential of the target nodes. Our findings suggest that in practice there is a trade-off that we need to make between optimizing for influence potential and for navigational boost. For the former, we need to aim at target nodes with a high degree ratio and for the latter at target nodes with a low degree ratio. 

\subsection{Combinations}
In the previous experiments we found that in some situations link insertion should be preferred over click bias (e.g., small fraction $\phi$ of target nodes), whereas sometimes the opposite represents an optimal approach (e.g., large $\phi$). For that reason we want now to shed more light onto combinations of both strategies, that is, we are interested in the navigational effects of simultaneously applying click bias and link insertion to varying extent. Figure~\ref{fig:combinations} depicts the results of this experiment. We find consistent best performing mixtures over all datasets. In particular, we observe that for small fractions $\phi$ of target nodes, exclusive link insertion outperforms any other combination (see Figure~\ref{fig:combinations:001}). For medium sized target nodes (i.e., $\phi=0.1$) we observe a shift of best performing combinations towards $\alpha=0.9$ for higher bias strengths (i.e., $b=5$ and $b=15$). This combination consist of $90\%$ click bias and $10\%$ link insertion. For combinations of large fractions of target nodes (i.e., $\phi=0.2$) and small bias strengths ($b=2$) the best performing combination is around $\alpha=0.5$ ($50\%$ click bias and $50\%$ link insertion) and further shifts towards $\alpha=0.9$ ($90\%$ click bias and $10\%$ link insertion) with an increased bias strength. 

These results confirm our insights from the previous experiments. Thus, click biases act as an amplifier and only work well if target nodes initially possess valuable incoming links. This is highly likely for larger and medium sized fractions of target nodes, and very unlikely for the case of smaller fractions of target nodes. On the other hand, link insertion diffuses a large portion of the energy of top nodes towards target nodes. Hence, it works especially well for combinations of small fractions of target nodes and datasets with a highly skewed stationary distribution.

\findingbox{For small fractions of target nodes with initially low energy, pure link insertion should be preferred over any other combination. However, with increasing bias strength and larger fraction of target nodes, combinations consisting of $90\%$ click bias and $10\%$ link insertion performs best.}

\noindent\textbf{Implications.} Smaller sets of webpages\xspace (i.e., small $\phi$) should focus on introducing new links to achieve the highest browsing guidance. The bigger the set of webpages\xspace and the used bias strength becomes, the more this preference shifts towards a combination of $0.9$, meaning that $90$\% of the modifications should be invested in increasing the transition probability of already existing links towards target nodes (e.g., highlighting in the user interface). The remaining $10$\% should be used to insert new links towards target nodes.

\subsection{Stationary vs. Transient User Behavior}
The random surfer which navigates forever (stationary behavior) may look like a rather unrealistic behavior of users. More realistically, a single user visits a website\xspace clicks a couple of times on various links and leaves the website\xspace again (transient behavior). However, our calculations of the stationary distribution show that, at least on the networks that we have investigated in this paper these two behaviors are quite similar to each other. 

The stationary distribution is calculated with the power-iteration method~\cite{golub2012matrix}. Thus, we initialize a probability vector representing an initial probability to find a random surfer on each particular node in the network. We initialize this vector using a uniform distribution.
Afterwards, we iterate by recalculating the probabilities for the next click of the random surfer. Thus, one iteration step of the power-iteration method can be interpreted as a step or a click performed by the random surfer moving from the current node to one of its neighbors. Hence, the number of iteration steps that are needed until there are no significant changes in the node probabilities, that is, the convergence rate of the power-iteration method, can be interpreted as the number of clicks needed to model the stationary user behavior. In other words the random surfer does not need to navigate forever---it only needs to navigate through the network until the point where the next click does not change the observed stationary distribution. 

In all our datasets, all networks that we generated and modified for these datasets, all combinations of fractions of target nodes $\phi$ and the bias strength $b$ our calculations converge within $8$ iterations. Thus, the stationary user behavior is in fact a behavior of users who navigate $8$ pages in a website\xspace at most. We believe that these $8$ clicks are within realistic boundaries for user behavior in the cases in which users decide to explore and browse a website\xspace. However, since many users leave a website\xspace immediately upon arrival or within only a single or a small number of clicks this still represents a limitation in our work. This limitation can be easily remedied by introducing a small teleportation probability of jumping to an arbitrary page without following the underlying network structure (i.e., calculating PageRank vector instead of the stationary distribution). We have already experimented with the calculations of PageRank and our first results are quite similar to results that we have presented in this paper. However, we plan to address this question in more details in our future work.

\section{Conclusions}
In this paper we have analyzed the effects of two link modification strategies used to influence the typical whereabouts of the random surfer. We investigated how an induced click bias towards a set of webpages\xspace changes the stationary distribution (i.e., energy) of those pages. Additionally, we compared those effects with the consequences of altering the network structure by inserting new links. We find that both strategies have a high potential to modify the stationary distribution and that for certain situations there exist constantly high performing link modification strategy. In particular, click biases work well on sets of webpages\xspace containing already highly visible webpages\xspace, whereas link insertion should be preferred for sets of webpages\xspace consisting of pages with low visibility. Further, we showed that a simple structural property of target nodes, namely degree ratio, provides a valuable basis for the estimation of the effects of both link modification strategies. Administrators of websites\xspace can use our approach and our open source framework to determine the best strategy for their settings without having to implement and test all the different strategies (e.g., altering link position, highlighting, or creating new links). %

In future work our analysis can be extended to investigate additional empirical as well as synthetic datasets to broaden the understanding of consequences of manipulating the link selection process in navigation or inserting new links. Furthermore, investigating the complex dynamics which arise if we induce two competing link modifications into one network at the same time is an interesting avenue for future work.

\balance
\bibliographystyle{abbrv}

\begin{thebibliography}{10}

\bibitem{boundsoftopicsenspagerank}
S.~Al-Saffar and G.~Heileman.
\newblock Experimental bounds on the usefulness of personalized and
  topic-sensitive pagerank.
\newblock In {\em Web Intelligence, IEEE/WIC/ACM International Conference on},
  pages 671--675. IEEE, 2007.

\bibitem{bian2011online}
L.~Bian and H.~Holtzman.
\newblock Online friend recommendation through personality matching and
  collaborative filtering.
\newblock {\em Proc. of UBICOMM}, pages 230--235, 2011.

\bibitem{insidepr}
M.~Bianchini, M.~Gori, and F.~Scarselli.
\newblock Inside pagerank.
\newblock {\em ACM Trans. Internet Technol.}, 5(1):92--128, Feb. 2005.

\bibitem{brin}
S.~Brin and L.~Page.
\newblock Reprint of: The anatomy of a large-scale hypertextual web search
  engine.
\newblock {\em Computer networks}, 56(18):3825--3833, 2012.

\bibitem{buscher2009you}
G.~Buscher, E.~Cutrell, and M.~R. Morris.
\newblock What do you see when you're surfing?: using eye tracking to predict
  salient regions of web pages.
\newblock In {\em Proceedings of the SIGCHI conference on human factors in
  computing systems}, pages 21--30. ACM, 2009.

\bibitem{dimi_www_poster}
D.~Dimitrov, P.~Singer, F.~Lemmerich, and M.~Strohmaier.
\newblock {Visual Positions of Links and Clicks on Wikipedia}.
\newblock In {\em Proceedings of the 25th International Conference on World
  Wide Web}, WWW '16 Companion, New York, NY, USA, 2016. ACM.

\bibitem{ding2002pagerank}
C.~Ding, X.~He, P.~Husbands, H.~Zha, and H.~D. Simon.
\newblock Pagerank, hits and a unified framework for link analysis.
\newblock In {\em Proceedings of the 25th annual international ACM SIGIR
  conference on Research and development in information retrieval}, pages
  353--354. ACM, 2002.

\bibitem{ding2004link}
C.~H. Ding, H.~Zha, X.~He, P.~Husbands, and H.~D. Simon.
\newblock Link analysis: hubs and authorities on the world wide web.
\newblock {\em SIAM review}, 46(2):256--268, 2004.

\bibitem{geigl_iknow}
F.~Geigl, D.~Lamprecht, R.~Hofmann-Wellenhof, S.~Walk, M.~Strohmaier, and
  D.~Helic.
\newblock Random surfers on a web encyclopedia.
\newblock In {\em Proceedings of the 15th International Conference on Knowledge
  Technologies and Data-driven Business}, i-KNOW '15, pages 5:1--5:8, New York,
  NY, USA, 2015. ACM.

\bibitem{Gleich2010}
D.~F. Gleich, P.~G. Constantine, A.~D. Flaxman, and A.~Gunawardana.
\newblock Tracking the random surfer: empirically measured teleportation
  parameters in pagerank.
\newblock In {\em Proceedings of the 19th international conference on World
  wide web}, pages 381--390. ACM, 2010.

\bibitem{golub2012matrix}
G.~H. Golub and C.~F. Van~Loan.
\newblock {\em Matrix computations}, volume~3.
\newblock JHU Press, 2012.

\bibitem{Gyongyi2004}
Z.~Gy{\"o}ngyi, H.~Garcia-Molina, and J.~Pedersen.
\newblock Combating web spam with trustrank.
\newblock In {\em Proceedings of the Thirtieth international conference on Very
  large data bases-Volume 30}, pages 576--587. VLDB Endowment, 2004.

\bibitem{Haveliwala2002}
T.~H. Haveliwala.
\newblock Topic-sensitive pagerank.
\newblock In {\em Proceedings of the 11th international conference on World
  Wide Web}, pages 517--526. ACM, 2002.

\bibitem{haveliwala2003topic}
T.~H. Haveliwala.
\newblock Topic-sensitive pagerank: A context-sensitive ranking algorithm for
  web search.
\newblock {\em Knowledge and Data Engineering, IEEE Transactions on},
  15(4):784--796, 2003.

\bibitem{Helic2013}
D.~Helic, M.~Strohmaier, M.~Granitzer, and R.~Scherer.
\newblock Models of human navigation in information networks based on
  decentralized search.
\newblock In {\em Proceedings of the 24th ACM Conference on Hypertext and
  Social Media}, pages 89--98. ACM, 2013.

\bibitem{hoggDisentangling}
T.~Hogg and K.~Lerman.
\newblock Disentagling the effects of social signals.
\newblock {\em Human Computation Journal}, 2(2):189--208, 2015.

\bibitem{KleinbergHITS}
J.~M. Kleinberg.
\newblock Authoritative sources in a hyperlinked environment.
\newblock {\em Journal of the ACM (JACM)}, 46(5):604--632, 1999.

\bibitem{deepinsidepr}
A.~N. Langville and C.~D. Meyer.
\newblock Deeper inside pagerank.
\newblock {\em Internet Mathematics}, 1(3):335--380, 2004.

\bibitem{lerman_pos_bias}
K.~Lerman and T.~Hogg.
\newblock Leveraging position bias to improve peer recommendation.
\newblock {\em PLoS ONE}, 9(6):e98914, 06 2014.

\bibitem{Li2009}
N.~Li and G.~Chen.
\newblock Multi-layered friendship modeling for location-based mobile social
  networks.
\newblock In {\em Mobile and Ubiquitous Systems: Networking Services,
  MobiQuitous, 2009. MobiQuitous '09. 6th Annual International}, pages 1--10,
  July 2009.

\bibitem{lovasz1993random}
L.~Lov{\'a}sz.
\newblock Random walks on graphs: A survey.
\newblock {\em Combinatorics, Paul erdos is eighty}, 2(1):1--46, 1993.

\bibitem{Moricz2010}
M.~Moricz, Y.~Dosbayev, and M.~Berlyant.
\newblock Pymk: Friend recommendation at myspace.
\newblock In {\em Proceedings of the 2010 ACM SIGMOD International Conference
  on Management of Data}, SIGMOD '10, pages 999--1002, New York, NY, USA, 2010.
  ACM.

\bibitem{murphy2006primacy}
J.~Murphy, C.~Hofacker, and R.~Mizerski.
\newblock Primacy and recency effects on clicking behavior.
\newblock {\em Journal of Computer-Mediated Communication}, 11(2):522--535,
  2006.

\bibitem{ogras2006s}
U.~Y. Ogras and R.~Marculescu.
\newblock " it's a small world after all": Noc performance optimization via
  long-range link insertion.
\newblock {\em Very Large Scale Integration (VLSI) Systems, IEEE Transactions
  on}, 14(7):693--706, 2006.

\bibitem{PRordertheweb}
L.~Page, S.~Brin, R.~Motwani, and T.~Winograd.
\newblock The pagerank citation ranking: bringing order to the web.
\newblock 1999.

\bibitem{pandurangan2002using}
G.~Pandurangan, P.~Raghavan, and E.~Upfal.
\newblock Using pagerank to characterize web structure.
\newblock In {\em Computing and Combinatorics}, pages 330--339. Springer, 2002.

\bibitem{richardson2001intelligent}
M.~Richardson and P.~Domingos.
\newblock The intelligent surfer: Probabilistic combination of link and content
  information in pagerank.
\newblock In {\em NIPS}, pages 1441--1448, 2001.

\bibitem{silva2010}
N.~Silva, I.-R. Tsang, G.~Cavalcanti, and I.-J. Tsang.
\newblock A graph-based friend recommendation system using genetic algorithm.
\newblock In {\em Evolutionary Computation (CEC), 2010 IEEE Congress on}, pages
  1--7, July 2010.

\bibitem{west2012automatic}
R.~West and J.~Leskovec.
\newblock Automatic versus human navigation in information networks.
\newblock In {\em ICWSM}, 2012.

\bibitem{woess1994random}
W.~Woess.
\newblock Random walks on infinite graphs and groups-a survey on selected
  topics.
\newblock {\em Bulletin of the London Mathematical Society}, 26(1):1--60, 1994.

\bibitem{Xie2010}
X.~Xie.
\newblock Potential friend recommendation in online social network.
\newblock In {\em Green Computing and Communications (GreenCom), 2010 IEEE/ACM
  Int'l Conference on Int'l Conference on Cyber, Physical and Social Computing
  (CPSCom)}, pages 831--835, Dec 2010.

\end{thebibliography}

\end{document}